\documentclass[12pt]{iopart}
\usepackage{import}
\usepackage{caption}
\usepackage{graphicx,comment}
\usepackage{amsmath}
\usepackage{amssymb}
\usepackage{amstext}
\usepackage{amsfonts}
\input{Qcircuit}
\usepackage{bm}
\usepackage[normalem]{ulem}
\usepackage{color}
\usepackage{float}
\usepackage{hyperref}
\usepackage{multirow}
\usepackage[table,xcdraw]{xcolor}
\usepackage{tabularx}
\usepackage{wrapfig}
\hypersetup{
     colorlinks   = true,
     citecolor    = blue
}
\usepackage{lineno}
\usepackage{color, colortbl}
{\rm }

\def\S{\Sigma}

\def\be{\begin{equation}}
 \def\ee{\end{equation}}
 \def\bea{\begin{eqnarray}}
 \def\eea{\end{eqnarray}}
\begin{document}
%\title {Measuring the Stability of NISQ Hardware Using Cycle Benchmarking}
%\title{Using a 1+1 Quantum Field Theory to Measure the Stability of NISQ Gate-Based Hardware}

\title[NISQ Gate-Based Qubit Stability Using Cycle Benchmarking]{Measuring NISQ Gate-Based Qubit Stability Using a 1+1 Field Theory and Cycle Benchmarking}
%\title{Measuring NISQ %Gate-Based Qubit Stability %Using a 1+1 Field Theory %and Cycle Benchmarking}
%\title{Measuring the Stability of NISQ Gate-Based Hardware Using Cycle Benchmarking}

\author{K\"ubra Yeter-Aydeniz
\footnote{current address:  Emerging Engineering and Physical Sciences Department. \\
The MITRE Corporation, 7515 Colshire Drive, McLean, VA 22102-7539, USA.}}
\address{Physics Division and Computational Sciences and Engineering Division \\
Oak Ridge National Laboratory,  Oak Ridge, TN 37831, USA}

\author{Zachary Parks}
\address{Department of Computer Science, North Carolina State University \\
Raleigh, North Carolina 27695}

\author{Aadithya Nair}
\address{Department of Computer Science, North Carolina State University \\
Raleigh, North Carolina 27695}

\author{Erik Gustafson \footnote{current address: Theoretical Physics Division, Fermi National Accelerator Lab, Batavia, IL 60510}}
\address{Department of Physics and Astronomy, The University of Iowa, Iowa City, IA 52242}

\author{Alexander F. Kemper}
\address{Department of Physics, North Carolina State University, Raleigh, North Carolina 27695}

\author{Raphael C.\ Pooser}
\address{Computational Sciences and Engineering Division, Oak Ridge National Laboratory \\
Oak Ridge, TN 37831, USA}

\author{Yannick Meurice}
\address{Department of Physics and Astronomy, The University of Iowa, Iowa City, IA 52242}

\author{Patrick Dreher \footnote{Corresponding Author: email: padreher@ncsu.edu }}
\address{Electrical and Computer Engineering and Department of Physics \\
North Carolina State University, Raleigh, NC  27695, USA}

\date{\today}

\begin{abstract}
 Some of the most problematic issues that limit the implementation of applications on Noisy Intermediate Scale Quantum (NISQ) machines are the adverse impacts of both incoherent and coherent errors. We conducted an in-depth study of coherent errors on a quantum hardware platform using a transverse field Ising model Hamiltonian as a sample user application. We report here on the results from these computations using several error mitigation protocols that profile these errors and provide an indication of the qubit stability. Through a detailed set of measurements we identify inter-day and intra-day qubit calibration drift and the impacts of quantum circuit placement on groups of qubits in different physical locations on the processor. This paper also discusses how these measurements can provide a better understanding of these types of errors and how they may improve efforts to validate the accuracy of quantum computations.
\footnote{This manuscript has been authored by UT-Battelle, LLC, under Contract No. DE-AC0500OR22725 with the U.S. Department of Energy. The United States Government retains and the publisher, by accepting the article for publication, acknowledges that the United States Government retains a non-exclusive, paid-up, irrevocable, world-wide license to publish or reproduce the published form of this manuscript, or allow others to do so, for the United States Government purposes. The Department of Energy will provide public access to these results of federally sponsored research in accordance with the DOE Public Access Plan.}

\end{abstract}

\maketitle

\section{Introduction}
\label{sec:intro}

Today researchers and application developers have access to the first generation of Noisy Intermediate Scale Quantum (NISQ)~\cite{Preskill2018quantumcomputingin} quantum computing (QC) hardware platforms.  This has opened opportunities for users to begin exploring how to reformulate existing algorithms designed for digital computers onto quantum computing hardware platforms.  These machines also offer an operational environment where new algorithms specially optimized for quantum computers can be developed.  These new capabilities can now provide the quantum computing user community with an ``on-ramp" into the world of quantum computing to explore problems in the near future that, up to this point, have been inaccessible using even the most powerful digital high performance computers.  

Although current quantum computing hardware capabilities now allow some basic applications to be successfully implemented, at the present time QC platforms utilizing various hardware architectures can only offer a range from a few qubits up to a few hundred noisy qubits over the next three to five years.  These noisy qubits degrade the fidelity of the quantum computations. This limitation on the number of qubits excludes any possibility of constructing ``logical qubits" to compensate and correct for the errors introduced by the noise in the quantum system.  These NISQ machines can only maintain a coherent quantum state for short periods of time. These times are usually expressed in standard metrics of T1 (relaxation time) and T2 (dephasing time) and two qubit CNOT error rates.

Understanding these different types of errors and developing methods to mitigate them is a critical area of research essential for advancing Quantum Information Science.  The types of errors that may occur when running algorithms on quantum hardware can be categorized as incoherent or coherent errors.  

Incoherent errors are due to uncontrolled interactions between qubits and the environment that result in the decoherence of the overall quantum state.  One particular example of a technique to mitigate decoherence is to scale up the level of noise by introducing pairs of CNOT gates, measure the output in each instance, and then extrapolating the signal output back to a zero-noise limit.  This has been extensively studied over the past few years. ~\cite{PhysRevLett.119.180509, PhysRevX.7.021050, Giurgica_Tiron_2020}. 

Coherent errors are mainly caused by systematic errors in the control of the qubits. This type of error results from over or under rotation in qubit control pulses that can result in calibration errors, drift in the qubit properties, or crosstalk where a nearby qubit is impacted by the coherent rotations directed at a different qubit.  These errors are quantified through norm-based error-metrics such as the diamond distance \cite{Wallman2016} and total variation distance \cite{hashim2021randomized} that are defined through a rigorous mathematical representation of unitary operators rotated through an angle $\theta$ relative to an intended target state.  

Developing a full characterization of these errors on quantum hardware platforms is a challenge because doing so requires quantum process tomography~\cite{2008qpt,2012stevenqpt,ref:Chuang1997, PhysRevLett.101.220501, 2009qptBrandehorst} or gate set tomography~\cite{ref:Merkel2012,ref:Blume-Kohout2013,ref:Blume-Kohout2017}, both of which require resources that grow exponentially with the number of qubits.  Alternatives to such an exhaustive option are process fidelity measurements such as Randomized Benchmarking (RB)~\cite{Emerson_2005,ref:Dankert2009,ref:Magesan2011}, Cycle Benchmarking (CB)~\cite{Erhard2019} and Randomized Compiling (RC)~\cite{Wallman2016}. Both CB and RC are better able to characterize the cross talk errors in the quantum circuit \cite{PhysRevLett.122.110501, Preskill2018quantumcomputingin, PhysRevLett.122.200502,2020crossresonance, Piltz_2014, Reagor_2018,2020crosstalk, 2011gaterigetti,tripathi2021suppression,Bialczak_2010, Zhao_2021, long2021universal, PhysRevA.79.062314}.  

Using the IBM Quantum \texttt{ibmq\_boeblingen} processor, we did an initial examination of the processor's re-calibration data and backend properties.  There were many seemingly large random fluctuations occurring in several of the single qubit gate properties from day-to-day.  Some of these results are illustrated in Table~\ref{table:Single-qubit-gate-error-layout2-on-1-24-and-1-29} in \ref{Appendix:Tables}.  For example, qubit 6 shows some extreme fluctuations in the measurements for  $T_{1}, T_{2}$, the IBM Quantum backend readout error and the values for the basis gates $U_{2}$ and $U_{3}$ between the morning of January $24^{th}$ and $29^{th}$.  Other single qubits also showed similar fluctuations in their backend properties.

Because it is known that, in general, two qubit gates contribute a substantially larger error on QC hardware platforms than do single qubit gates, our group conducted an in-depth stability analysis of these two-qubit gates on \texttt{ibmq\_boeblingen} by measuring the process infidelities using CB and comparing them to the results obtained from RB measurements based on the IBM quantum processor qubit re-calibrations.  Because of its wide applicability in quantum field theories and many-body interactions~\cite{Yeter_Aydeniz_2021,Lamm:2018siq,PhysRevA.79.062314,CerveraLierta2018exactisingmodel, Kandala_2019,PhysRevLett.119.180509, somma2016quantum, Johanning_2009, Smith_2019,PhysRevX.7.021050,vovrosh2020confinement,Salath__2015,2017Natur.551..579B,2017Natur.551..601Z,Labuhn_2016,chertkov2021holographic,Kandala:2017aa,PhysRevLett.101.220501,PhysRevA.87.032341,PhysRevA.95.052339,Schauss_2018,10.1145/3408039,2017Natur.551..579B,keesling2019,2021arXiv210809197K} we used two-qubit gates in the Transverse Field Ising Model (TFIM) for our study of the two-qubit gate error properties.

This paper is organized as follows. In Sec.~\ref{sec:model} we describe the Transverse Field Ising Model used to study the error characterization in an IBM superconducting transmon platform.  Sec.~\ref{sec:methodology} discusses the methodology and conditions that were implemented using this model and the types of data that were collected.  Sec.~\ref{sec:measurements} summarizes the results of the computations using this data that illustrate inter-day and intra-day quantum hardware processor calibration drift and other measures and tests of stability of the TFIM circuits.  Using the results from these measurements, a discussion of some of the implications of these results is presented.  Finally, Sec.~\ref{sec:summary_conc} summarizes our observations and and discusses next steps.

\section{Physics Model}
\label{sec:model}

We study the transverse field Ising model (TFIM) with open boundary conditions that has the system Hamiltonian 
\be
H=-J\sum_{i=1}^{N_s-1} \hat{X}_i \hat{X}_{i+1} - h_T \sum_{i=1}^{N_s} \hat{Z}_i. \label{eq:H_OBC}
\ee

The operators $\hat{X}_i$ and $\hat{Z}_i$ correspond to the Pauli matrices $\hat{\sigma}^x$ and $\hat{\sigma}^z$ respectively. $N_s$ corresponds to the number of sites in the model. $J$ is the nearest neighbor (hopping) coupling and controls the movement of the spins and creation of spin pairs, while $h_T$ is the on-site energy. This model has been used 
in a variety of contexts related to quantum computing~\cite{Yeter_Aydeniz_2021,Lamm:2018siq,PhysRevA.79.062314,CerveraLierta2018exactisingmodel, Kandala_2019,PhysRevLett.119.180509, somma2016quantum, Johanning_2009, Smith_2019,PhysRevX.7.021050,vovrosh2020confinement,Salath__2015,2017Natur.551..579B,2017Natur.551..601Z,Labuhn_2016,chertkov2021holographic,Kandala:2017aa,PhysRevLett.101.220501,PhysRevA.87.032341,PhysRevA.95.052339,Schauss_2018,10.1145/3408039,keesling2019,2021arXiv210809197K}.

In the following, we will study the case with $N_s=4$, $J=0.02$ and $h_T=1$. This choice of parameters provides a simple particle picture and has been used in recent quantum computing studies 
\cite{Gustafson:2021imb,Gustafson2021,GustafsonIsing}.

The system can be evolved in time using the complex exponential of the Hamiltonian:
\begin{equation}
\label{eqtimeevolveexact}
\hat{U}(t) = e^{- i t \hat{H}}.
\end{equation}
Following Refs.~\cite{Lloyd1073,GustafsonIsing,Gustafson2021},
the Trotter approximation 
is applied to the evolution operator 
with the explicit form:
\begin{equation}
\label{eqsuzuki}
\hat{U}(t;N) = \Big(\hat{U}_1(t / N; h_t) \hat{U}_2(t / N; J)\Big)^N + \mathcal{O}(t^2 / N)
\end{equation}
where $N$ is the number of Trotter steps to be implemented, $(\delta t=\frac{t}{N})$ is the Trotter step size. For the values of $J$ and $h_t$ chosen here, a time of approximately $t\sim 100$ ($Jt\sim 2$) is needed in order to observe changes of the occupations that can be interpreted as 
the motion of a particle across the size of the system. 

In our experiments, 
we used the exponentials of the 1- and 2-body operators in the Hamiltonian
\begin{equation}
\label{eqtfieldevo}
\hat{U}_1(\delta t; h_t) = e^{-i h_T \delta t \sum_{i = 1}^{4} \hat{Z}_i },
\end{equation}
and
\begin{equation}
\label{eqhoppingevo}
\hat{U}_2(\delta t; J) = e^{-i J \delta t \sum_{i = 1}^{3} \hat{X}_i \hat{X}_{i+1}}.
\end{equation}
We chose a Trotter step $\delta t=10$ which allows us to reach significant changes 
using five to ten steps. Notice that $\delta t=10$ is much larger than what would be required to 
control the error of one Trotter step with an accuracy proportional $\delta t^2$ or 
$\delta t^3$ for an improved Trotter approximation. Instead, it was noticed \cite{ybook,Gustafson2021} that for this nonlinear regime, the error grows linearly with a small coefficient. Recent work has shown that the standard error bounds are overly pessimistic \cite{PhysRevX.11.011020,2021arXiv210708032L}.

\begin{figure}[!htpb]
    % \centering
    % \includegraphics[width=2.2\columnwidth]{final_plot.pdf}
    %\includegraphics[width=0.49\textwidth]{circuit_designs_LK.pdf}
    \includegraphics[width=1.0\textwidth,
    clip=true,trim=18 15 20 10 0]{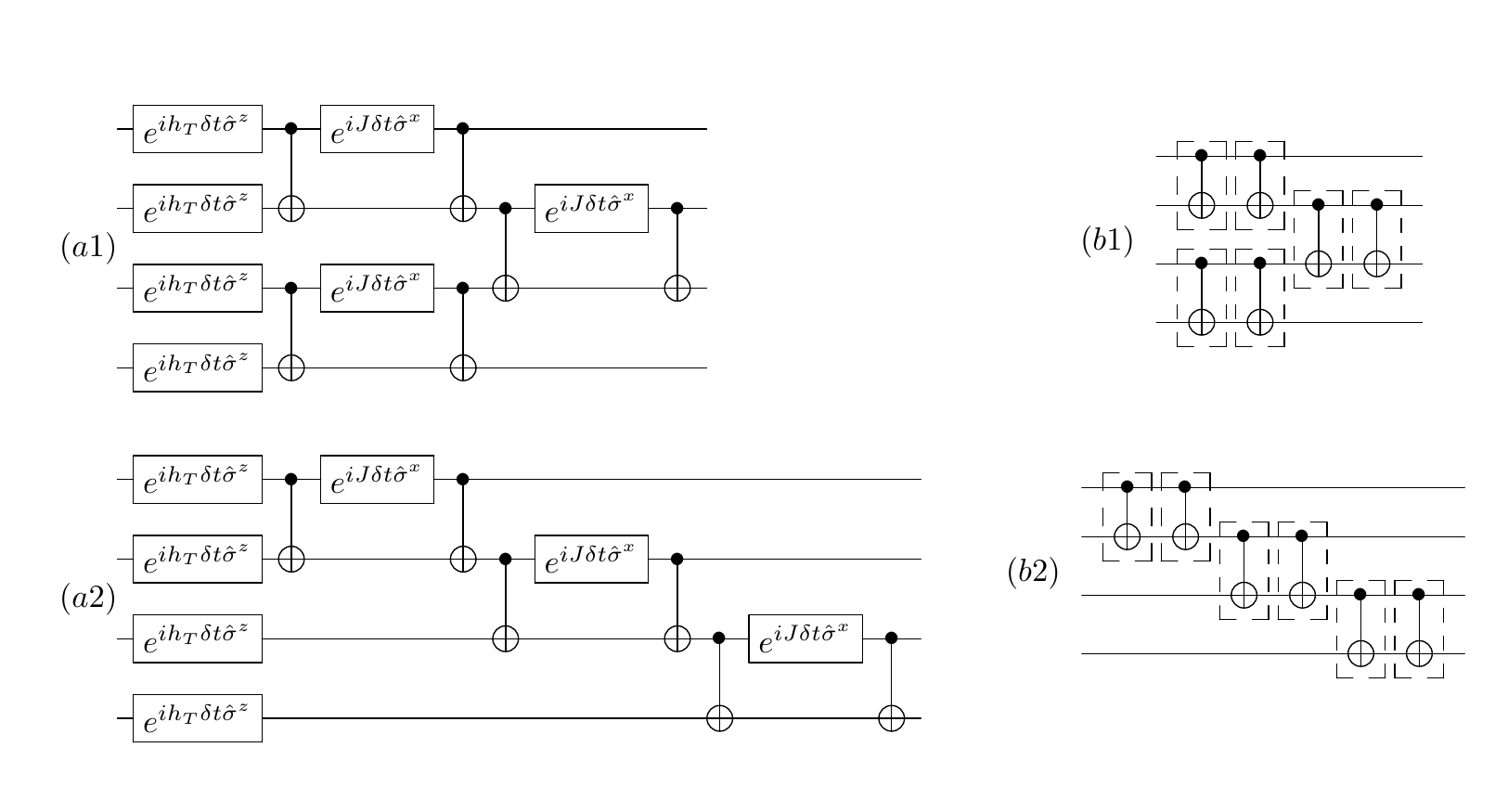}
    \caption
    {($a1$,$a2$): The quantum circuits for one Trotter step of the time evolution with the open boundary condition Ising model Hamiltonian. We define the quantum circuit in ($a1$) and ($a2$) as Circuit 1 and Circuit 2 respectively. \\
    ($b1$,$b2$): The CNOT hard cycles used at each Trotter step for Circuit 1 ($b1$) and Circuit 2 ($b2$).  The quantum circuits were drawn using the Q-circuit package \cite{QCircuit}.}
    \label{fig:circuit_design}
\end{figure}

The operators defined in Eqs. \ref{eqtfieldevo}  and \ref{eqhoppingevo} can be expressed as a combination of the two quantum circuit elements (1-qubit and 2-qubit gates) shown in Fig. \ref{fig:circuit_design}.

\section{Methodology}
\label{sec:methodology}

The Transverse Field Ising model (TFIM) Hamiltonian offered a model for studying the error characterization properties of multiple two-qubit CNOT gates. To model the TFIM Hamiltonian in Eq.~\eqref{eq:H_OBC}, we selected two quantum circuit diagrams of the TFIM (Fig.~\ref{fig:circuit_design}) that represent two choices or orderings for two-qubit components of this application (labelled as Circuit 1 and Circuit 2).  These two quantum circuits are equivalent combinations because the gates representing the different terms in the TFIM Hamiltonian commute with each other.  Both result in time evolution of a state with $\mathcal{U}=e^{-iH\delta t}$ using Trotterization where $\delta t$ is the time interval for one Trotter step.  Each circuit has three sets of CNOT gates.  We focused on measuring CNOT gate performance because the error rates for two qubit gates are substantially higher than single qubit errors. 

\begin{figure}[htpb]
    \begin{center}
    \includegraphics[width=0.49\textwidth,
     clip=true,trim=0 0 0 0]{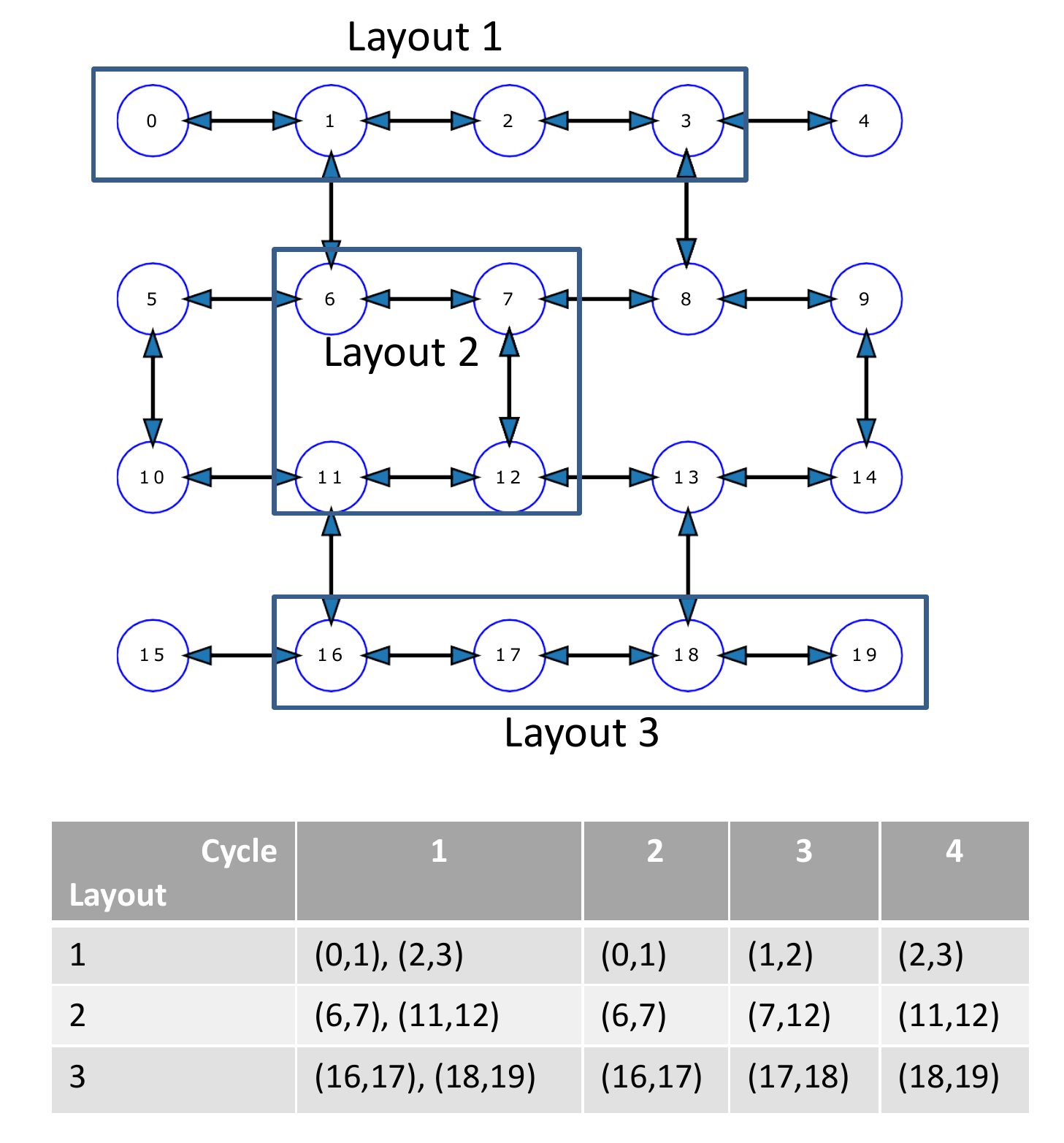}
     \end{center}
    \caption{Top: Qubit layout for the \texttt{ibmq\_boeblingen} quantum computer. Three different groups of qubits were selected to run the set of cycle benchmarking and TFIM computations described in this paper.  Layout 1 refers to qubits [0, 1, 2, 3], Layout 2 refers to qubits [6, 7, 12, 11] and Layout 3 refers to qubits [16, 17, 18, 19].  The exhaustive set of four different paired CNOTs (four different cycles) were used to calculate the process infidelities on each Layout on the \texttt{ibmq\_boeblingen}.processor}
    \label{fig:hardware_layout}
\end{figure}

We selected the 20-qubit \texttt{ibmq\_boeblingen} hardware platform (Fig.~\ref{fig:hardware_layout}) in order to study the stability of this application on a superconducting transmon device. Because the two-qubit gates in quantum circuits are a major source of the errors generated in a quantum computer, this project focused on studying the behavior of the combinations of CNOT gates under various conditions.  

The performance characteristics of the 2-qubit gates on the \texttt{ibmq\_boeblingen} processor representing the CNOT gates for the TFIM Hamiltonian were explored using Cycle Benchmarking (CB) \cite{Erhard2019} and the Quantum Capacity (QCAP).  We chose this protocol because CB is scalable for estimating the effect of all global and local error mechanisms that occur across multi-qubit quantum processors when a clock cycle of operations is applied to a specific quantum register. To calculate the process infidelity of the cycles of interest through CB and the QCAP bound we utilized the True-Q software by Keysight Technologies \footnote{https://trueq.quantumbenchmark.com/index.html}. True-Q is a software tool that provides methods to calibrate and optimize the performance of quantum devices.  \ref{Appendix_CBandQCAP} summarizes CB protocol in general and how it was implemented within the True-Q software in more detail.

Within the qubit layout on \texttt{ibmq\_boeblingen} we selected three separate groups of qubits as shown in Fig.~\ref{fig:hardware_layout} (Layouts 1, 2, and 3) to study the error characterization due to TFIM Trotterization.  Within each layout, an exhaustive combination of pairs of adjacent qubits was identified and labeled as Cycle 1 through 4 in order to measure the specific error characteristics of each qubit pair associated with each of the CNOT gates in the TFIM circuit. Each row in the table in Fig.~\ref{fig:hardware_layout} corresponds to the specific CNOT combinations for that specific layout. For example, Layout 1 measurements included all of the combination of two-qubit cycles [(0, 1) and (2, 3), (0, 1), (1, 2) and (2, 3)]. 

\begin{figure*}[ht]
    \begin{center}
    \includegraphics[width=0.8\textwidth]{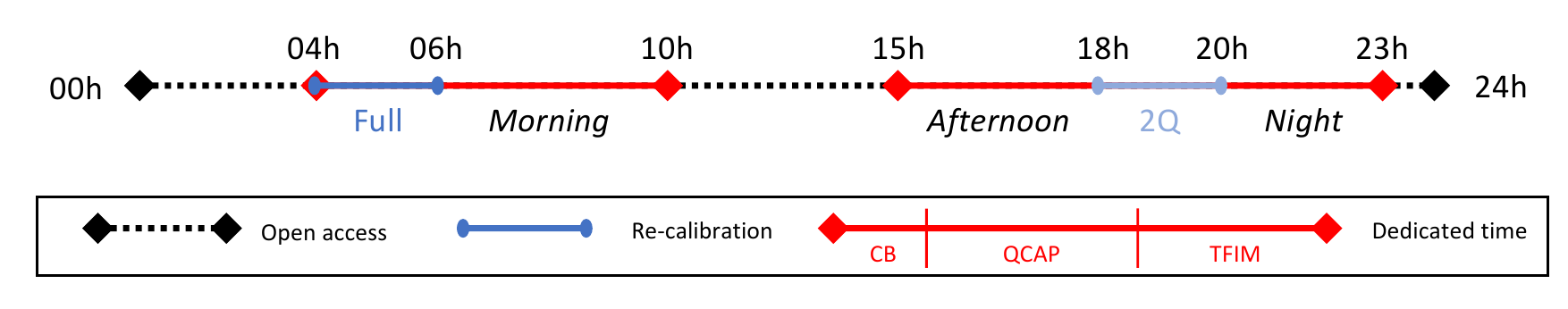}
    \end{center}
    \caption{Daily timeline for calibrations, computations and measurements. The red lines denote the dedicated reservation time periods. The blue lines denote the time periods when IBM performed a full 1 and 2 qubit re-calibration each morning and a two-qubit (2Q) only re-calibration each night.}
    \label{fig:timeline}
\end{figure*}

We ran a series of computations at three different preset time periods each day. Fig.~\ref{fig:timeline} shows the daily sequence of calibrations and measurements.  These computations and measurements were done over an 8 consecutive day time period in January 2021 while \texttt{ibmq\_boeblingen} was in a dedicated reservation mode of operation.  A dedicated reservation mode removes the hardware platform from general use and guarantees a user full dedicated access to the IBM hardware platform.  This mode of operation allowed the qubit calibrations, computations and measurements to run uninterrupted, undisturbed by other users accessing these qubits during this time.

Each day there was a block of dedicated reserved time on \texttt{ibmq\_boeblingen} from 4 am - 10 am and again from 3 pm - 11pm.  Each morning at 4 am IBM did a full re-calibration of all qubits and a 2 qubit re-calibration at 6 pm.  After the 4 am and 6 pm IBM re-calibrations were completed we recorded \texttt{ibmq\_boeblingen's} back-end properties.  \ref{IBM_re_calibrations} summarizes the daily IBM re-calibration schedule and procedures. 

At 4 a.m. each morning \texttt{ibmq\_boeblingen} was placed in a dedicated reservation mode.  Qubit re-calibrations, computations and ``morning" measurements were done from 6 am to 10 am.  The machine was then opened for general use from 10 am to 3 pm. At 3 pm the machine was closed to other users and again placed in dedicated reservation mode. A set of ``afternoon" measurements were made on the qubits from 3 pm to 6 pm.  The ``afternoon" measurements were deliberately scheduled when the machine was closed to all other users but prior to the night IBM re-calibration.  At 6 pm IBM did a two-qubit re-calibration.  After the re-calibration was finished another set of backend properties were recorded and then a full set of ``night" measurements were made from 8 pm - 11 pm.  

Running these experiments three different times on each day (morning, afternoon, night) for each of the three different Layouts with two different Circuits over an eight day consecutive time period gave inter-day and intra-day measurements of processor performance for each layout plus measurements based on qubit choice and circuit structure that could then be compared. 

The measurement were done using cycle benchmarking (CB) and randomized benchmarking (RB) to study the errors present in the circuits from Fig.~\ref{fig:circuit_design}.  Cycle Benchmarking (CB) was performed on Layouts 1, 2, and 3 using Circuits 1 and 2.  We computed the process infidelity and the quantum capacity bound (QCAP).  QCAP measurements were done on the three layouts only with Circuit 1.  After the CB computations were finished, the Trotterization of the TFIM Hamiltonian was run on each of the three different physical qubit layouts using only the Circuit 1 gate design so that a direct comparison could be made both to the previously published results in~\cite{GustafsonIsing} and the CB and QCAP measurements.  

\label{sec:inter-day-measure}

\section{Results}
\label{sec:measurements}
This section reports on the measurements obtained from the cycle benchmarking and TFIM computations on the IBM \texttt{ibmq\_boeblingen} processor during the 8 day time window in January 2021. The results reported here are unique in that the project had access to a very generous level of reserved run-time both in terms of dedicated access to the machine and length of contiguous dedicated time available for running detailed benchmarking computations.  Having this level of reserved time assured that the IBM re-calibrations were always performed on a quiescent machine and that the ``morning", ``afternoon" and ``night" CB, QCAP and TFIM computations fully ran without any other user's jobs being interleaved between the individual computations.  

These types of environmental factors assured an optimal set of run-time conditions for measuring the RB, CB and TFIM quantities.  \ref{Appendix_CBandQCAP} summarized the process infidelity and the QCAP bound computational procedure for analyzing the data.  Using these procedures, the analysis of the data focused on four areas:  inter-day and intra-day calibration drift of the qubits (Sec.~\ref{sec:inter-day-results} and \ref{sec:inter-day-analysis}), dependencies on qubit layouts for concurrent calculations (Sec.~\ref{sec:qubit choice}) and impacts from different circuit structure choices (Sec.~\ref{sec:circ1_vs_circ2_QCAP}).

\subsection{Inter-day and Intra-Day Qubit Hardware Performance}
\label{sec:inter-day-results}

\begin{figure*}[!htpb]
    % \centering
    % \includegraphics[width=2.2\columnwidth]{final_plot.pdf}
    \includegraphics[scale=0.5]{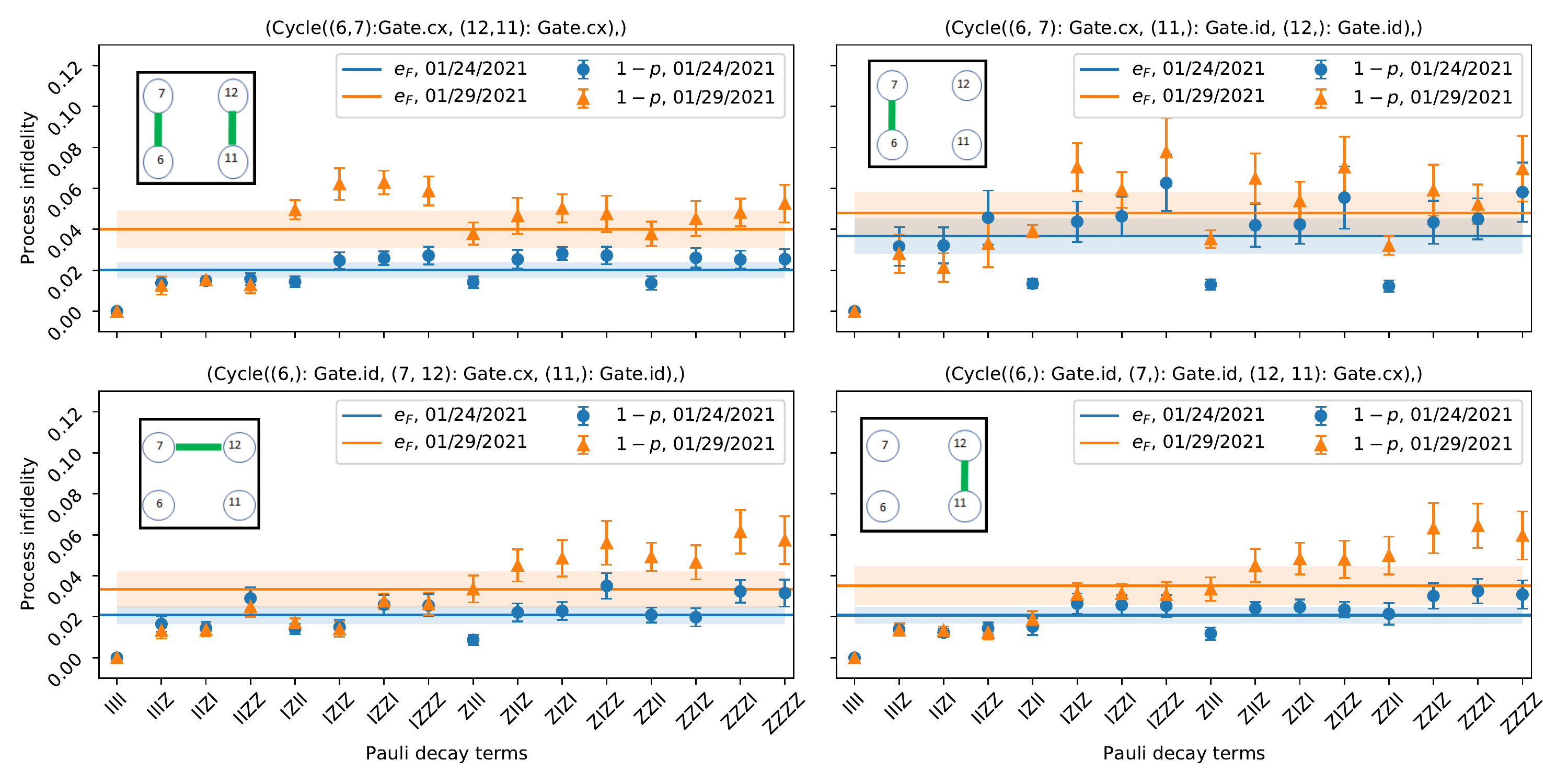}
    \caption{The Pauli infidelities for each hard cycle calculated for each Pauli decay term from morning run on qubits [6, 7, 12, 11] (layout 2) on days 01/24/2021 (blue lines and data points) and 01/29/2021 (orange lines and data points). The upper left graph is Cycle 1, the upper right is Cycle 2, lower left is Cycle 3 and the lower right is Cycle 4.  The total process infidelity for each of the 4 different cycles is graphed on each plot as a blue and orange solid lines along with the shaded error bands for that day ($24^{th}$ in blue and $29^{th}$ in orange).  The shaded regions show the error on the process infidelity and the error bars on the markers show the statistical errors on Pauli decay terms.}
    \label{fig:PauliInfidelities24_29_Story4}
\end{figure*}

Inter-day qubit drift was detected from analyzing data collected during the consecutive 8 day running period. An example illustrating this drift is seen by examining the data collected from Circuit 1, Layout 2 on January $24^{th}$ and January $29^{th}$. For both the January $24^{th}$ and January $29^{th}$ data, the Pauli infidelities for each hard cycle are calculated for each Pauli decay term from the morning run on Layout 2 (qubits [6, 7, 12,11]) and plotted in Fig~\ref{fig:PauliInfidelities24_29_Story4}.  

\begin{figure*}[htpb]
     \includegraphics[width=1.0\textwidth,
     clip=true,trim=0 280 0 0]{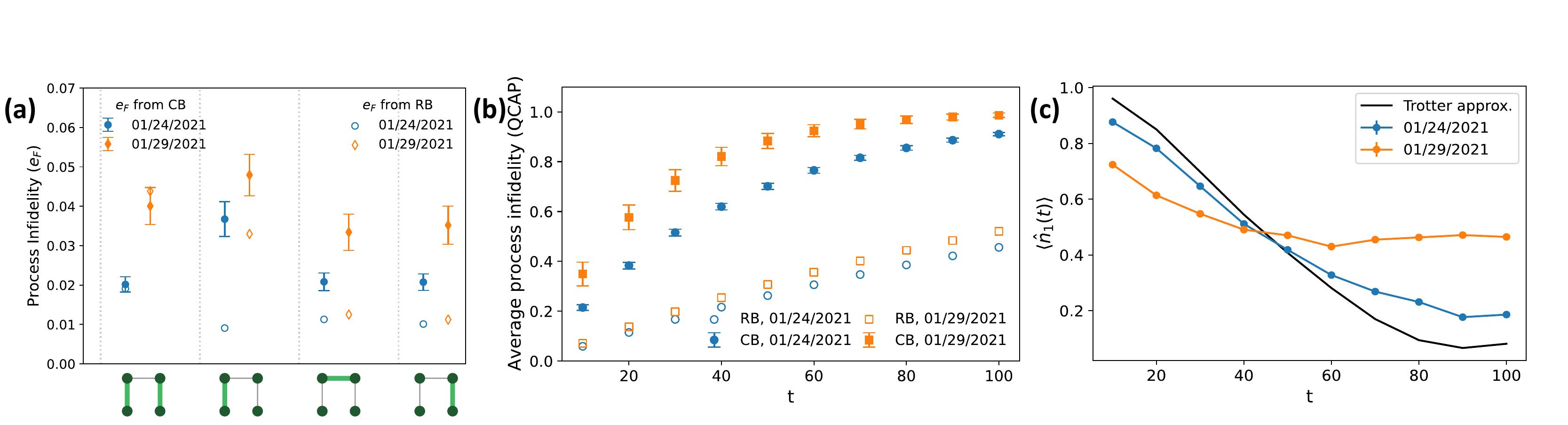}
    \caption{{\bf{Inter-day drift in error as characterized by RB and CB.}}
    % Left: individual gate set infidelities. Middle: Average process infidelities as a function
    % of number of Trotter steps, comparing two days and three times. Right: Corresponding
    % densities from the TFIM simulation. \KYA{Combined Story 4 plot}
    {\bf{(a)}} The process infidelities calculated using RB and CB such that the two-qubit CNOT cycles 1, 2, 3, and 4 are identified along the horizontal axis from left to right,
    {\bf{(b)}} The QCAP bound as a function of 
    %number Trotter steps (
    evolution time calculated using RB (QCAP$_{\textrm{RB}}$) and CB (QCAP$_{\textrm{CB}}$), 
    {\bf{(c)}} The particle number in site 1 calculated as a function of evolution time from morning run of layout 2 (qubits [6, 7, 12, 11]) on days 01/24/2021 and 01/29/2021 compared to exact Trotter approximation.
    }
    \label{fig:interdaydrift}
\end{figure*}

The RB two qubit error rates (r) that were recorded in the IBM backend properties after the completion of the IBM re-calibrations for those days were also included.  The average error rate, $r$ in standard RB assumes that the noise is gate-independent and $r\approx \epsilon$ where $\epsilon$ is the average gate-set infidelity. The average error rate  can be expressed in terms of RB decay rate $p$ as,
\begin{equation}
% \label{eq:avg process to gate infidelity conversion}
    r\equiv \frac{d-1}{d}(1-p) 
    \label{eq:err_rate}
\end{equation}
where $d=2^{n}$ (n is the number of qubits) \cite{Qi2019}. 
In this paper, we report the process infidelity, $e_F$, as a measure of infidelity for coherent errors. For completeness we convert average gate set fidelity, $r$, to the process infidelity, $e_F$ using,
\begin{equation}
    e_F=r \frac{d+1}{d}
    \label{eq:error_to_proc_infid}
\end{equation} 
from~\cite{NIELSEN2002249}.
The overall process infidelity from the CB calculations and the RB measured process infidelities were then plotted versus the exhaustive set of two-qubit CNOT cycles for Layout 2 measured on both the morning of January $24^{th}$ and January $29^{th}$ as shown in Fig.~\ref{fig:interdaydrift} {\bf{(a)}}.  The two-qubit error measurements are listed in Table~\ref{table:Two-qubit-gate-error-layout2-on-1-24-and-1-29} in \ref{Appendix:Tables}.

The QCAP bound as a function of number Trotter steps was then calculated using CB.  Both the QCAP$_{\textrm{CB}}$ and QCAP$_{\textrm{RB}}$ results were plotted in Fig.~\ref{fig:interdaydrift} {\bf{(b)}}.  After this QCAP data was collected the TFIM Trotterization on Circuit 1 was run while the \texttt{ibmq\_boeblingen} processor while the processor continued to be in dedicated mode.  Fig.~\ref{fig:interdaydrift}{\bf{(c)}} shows the occupation numbers on the first site $\langle \hat{n}_{1}(t)\rangle$ as a function of Trotter step/time calculated as a function of time from the morning runs of layout 2 (qubits [6, 7, 12, 11]) on January 24, 2021 and January 29, 2021 compared to exact Trotter approximation.

Intra-day qubit drift was also detected from data collected during the consecutive 8-day running period.  A similar procedure used to detect the inter-day qubit drift was also applied when analyzing data checking for intra-day drift.  Examples illustrating this drift are seen by examining the data collected from the morning, afternoon and night runs for both January $27^{th}$ and January $30^{th}$ using Layout 2.

The RB average process infidelity was recorded based the reported backend property of the two-qubit error rate after the IBM full processor qubit re-calibration.  The RB error rate was converted to an RB process infidelity in a similar procedure as was done for the inter-day RB data.  

\begin{figure*}[!htpb]
    % \centering
    % \includegraphics[width=2.2\columnwidth]{final_plot.pdf}
    \includegraphics[scale=0.52]{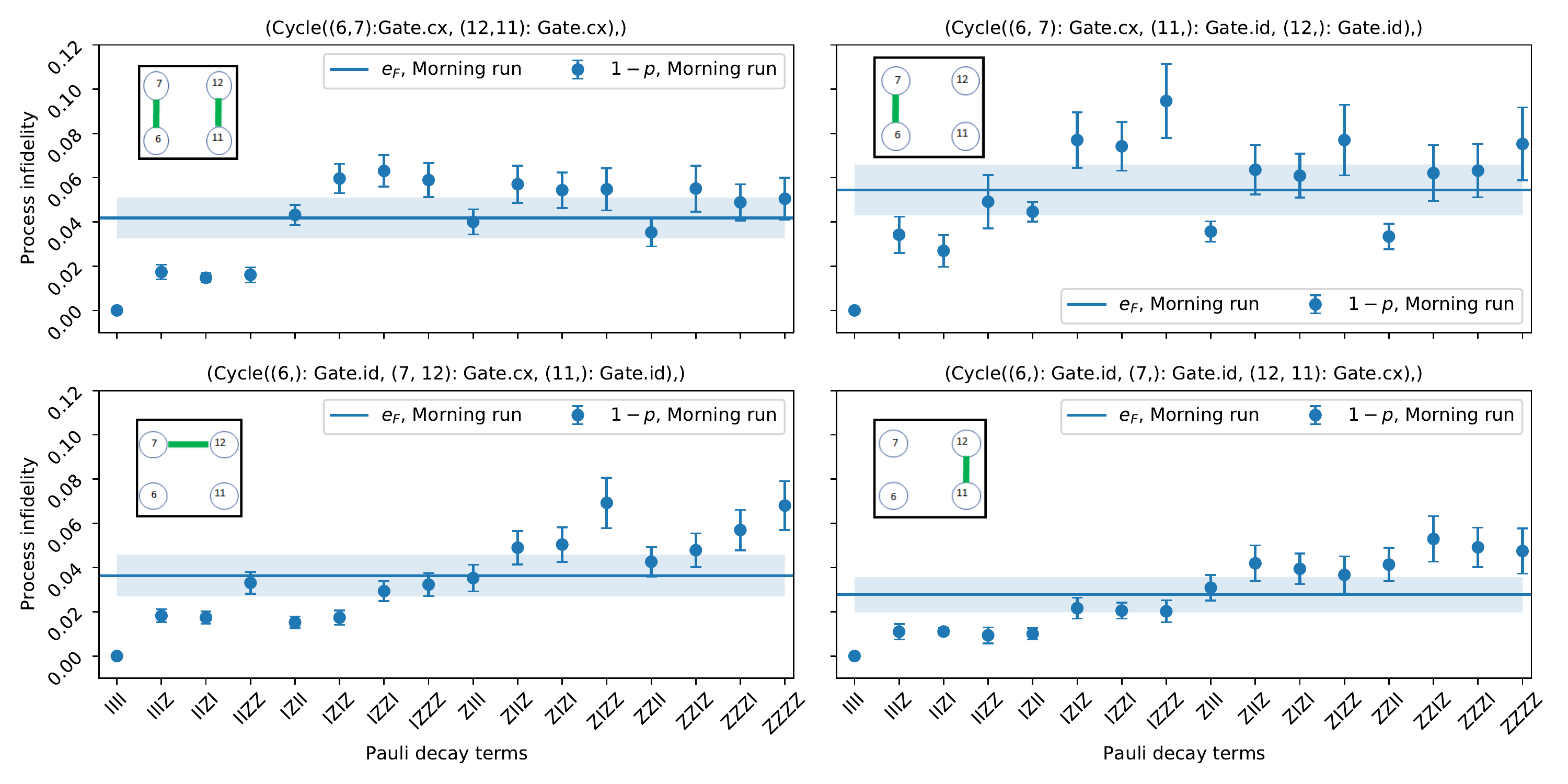}
    \caption{The Pauli infidelities for each hard cycle calculated for each Pauli decay term from morning run of layout 2 (qubits [6, 7, 12, 11]) on January 27, 2021. The error bars on each of the markers show the statistical errors on the Pauli decay terms.  The total process infidelity for each of the 4 different cycles are plotted as blue solid lines along with the shaded error bands.  The shaded regions show the error on the average process infidelity.}
    \label{fig:Proc_Infid_L2_Morning_1_27}
\end{figure*}

\begin{figure*}[!htpb]
    % \centering
    % \includegraphics[width=2.2\columnwidth]{final_plot.pdf}
    \includegraphics[scale=0.51]{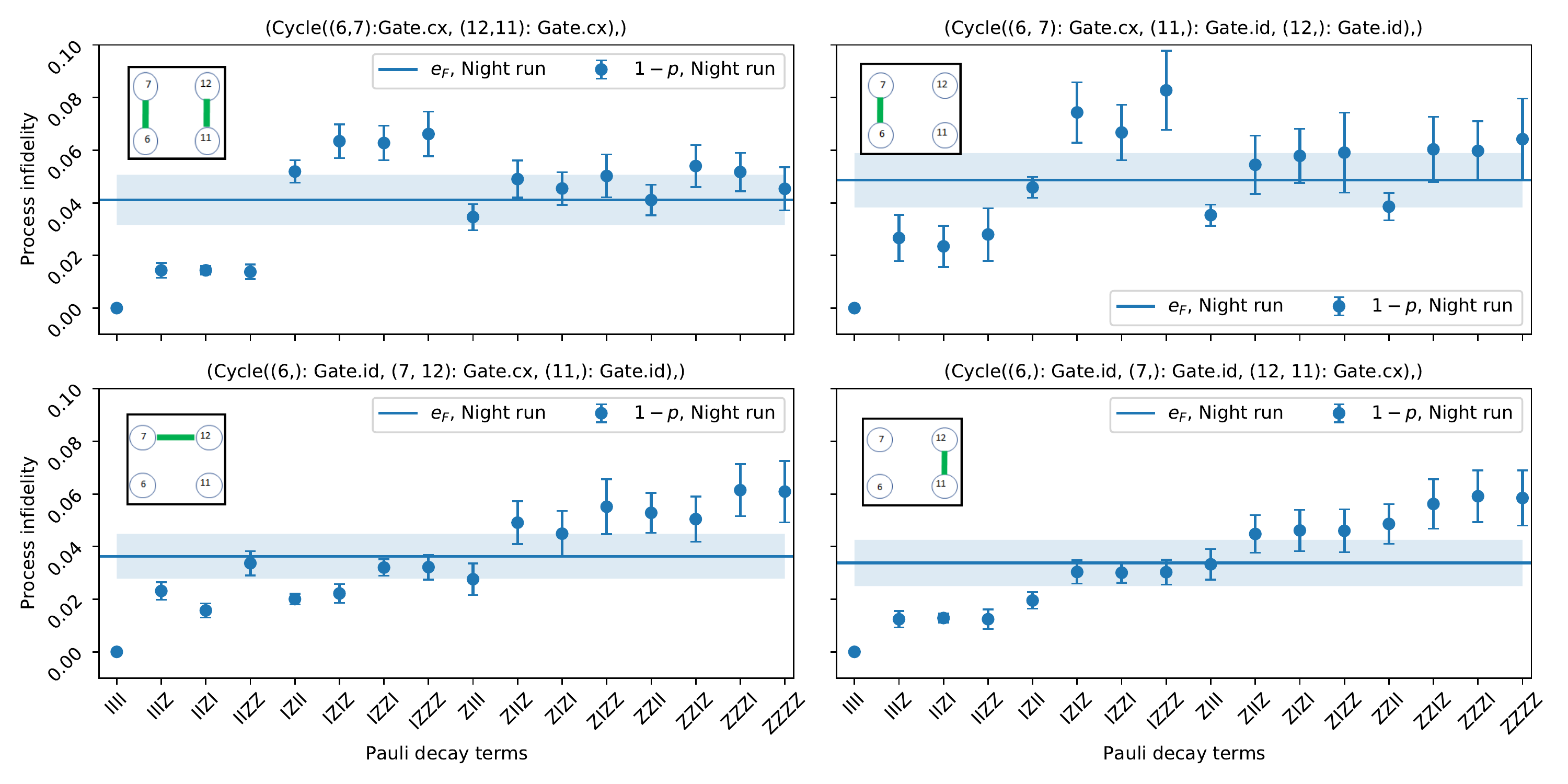}
    \caption{The Pauli infidelities for each hard cycle calculated for each Pauli decay term from night run of layout 2 (qubits [6, 7, 12, 11]) on day 01/30/2021. The error bars on each of the markers show the statistical errors on the Pauli decay terms.  The total process infidelity for each of the 4 different cycles are plotted as blue solid lines along with the shaded error bands.  The shaded regions show the error on the average process infidelity.}
    \label{fig:PauliInfidelities30Night}
\end{figure*}

The CB process infidelities for those days were also computed.  The individual process infidelities for each CNOT pair and the overall process infidelity are shown in  Fig.~\ref{fig:Proc_Infid_L2_Morning_1_27} thru   Fig.~\ref{fig:PauliInfidelities30Night}.

\begin{figure*}[ht]
    \centering
    \includegraphics[width=1.1\textwidth]{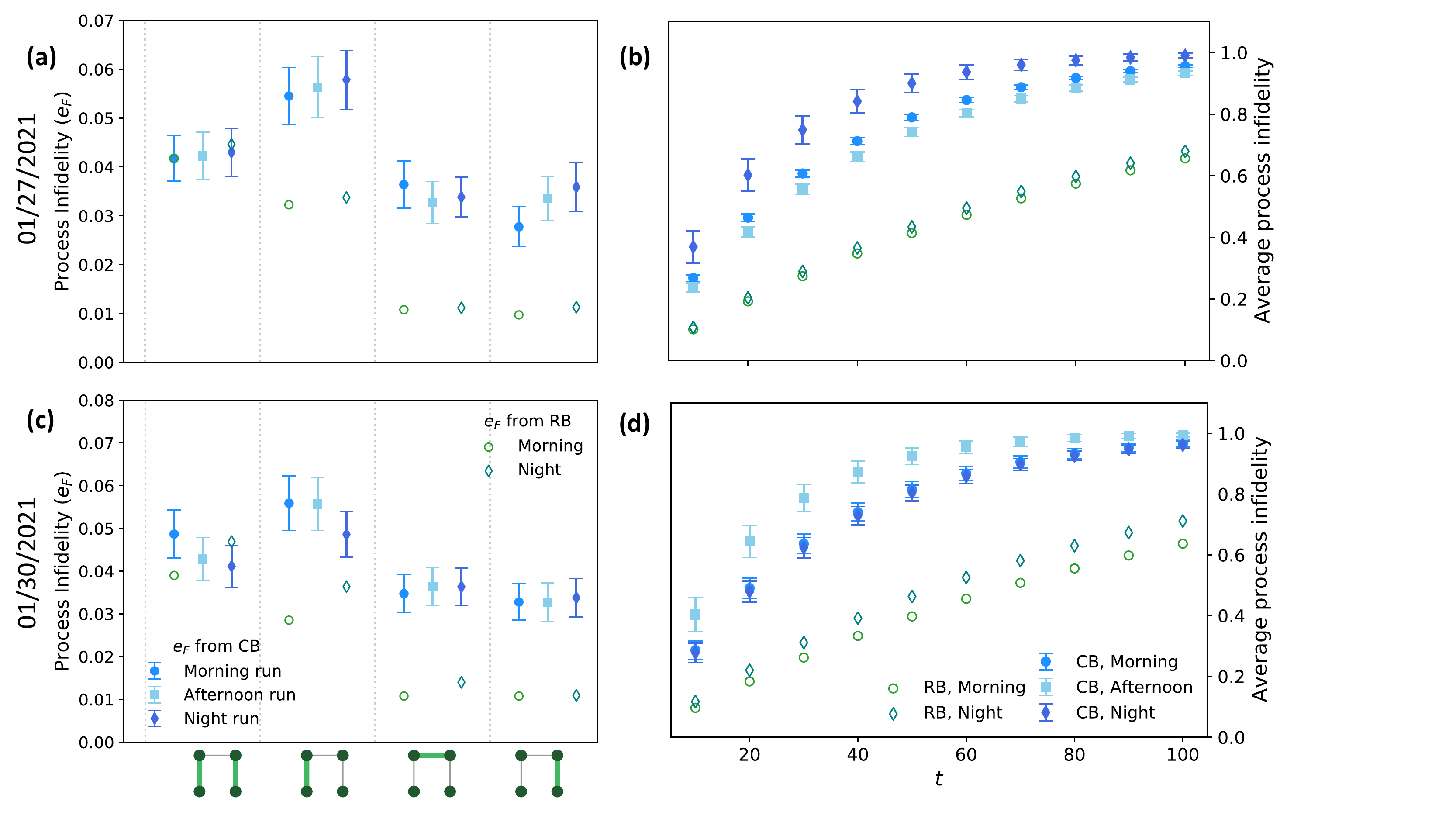}
    \caption{{\bf{Error characterization via CB vs. RB.}} 
    % Left: individual gate set infidelities. Right: Average process infidelities as a function
    % of number of Trotter steps, comparing two days and three times.
    The process infidelities for cycles  1, 2, 3, and 4 calculated using RB and cycle CB from morning, afternoon and night runs of layout 2 [qubits 6, 7, 12, 11] on {\bf{(a)}} January 27, 2021, {\bf{(c)}} January 30, 2021. The QCAP bound as a function of evolution time calculated using RB (QCAP$_{\textrm{RB}}$) and CB (QCAP$_{\textrm{CB}}$) from morning, afternoon and night runs of layout 2 [qubits 6, 7, 12, 11] on  {\bf{(b)}} January 27, 2021, {\bf{(d)}} January 30, 2021.
    }
    \label{fig:intraday_CB_vs_RB}
\end{figure*}

The overall process infidelity from the CB calculations and the RB measured process infidelities were plotted versus the exhaustive set of two-qubit CNOT cycles for Layout 2 measured on the morning, afternoon and night of January $27^{th}$ and January $30^{th}$.  Figs.~\ref{fig:intraday_CB_vs_RB} {\bf{(a)}} shows the CB versus RB process infidelity data and analysis and Fig.~\ref{fig:intraday_CB_vs_RB} {\bf{(b)}} shows the graph of the QCAP$_{\textrm{CB}}$ and the QCAP$_{\textrm{RB}}$ values versus evolution time for the January $27^{th}$ data.  Similarly Figs.~\ref{fig:intraday_CB_vs_RB} {\bf{(c)}} shows the CB versus RB process infidelity results and Figs.~\ref{fig:intraday_CB_vs_RB} {\bf{(d)}} shows the graph of the QCAP$_{\textrm{CB}}$ and the QCAP$_{RB}$ values versus evolution time for the January $30^{th}$ data.

\begin{figure*}[!htpb]
    % \centering
    % \includegraphics[width=2.2\columnwidth]{final_plot.pdf}
    \includegraphics[scale=0.51]{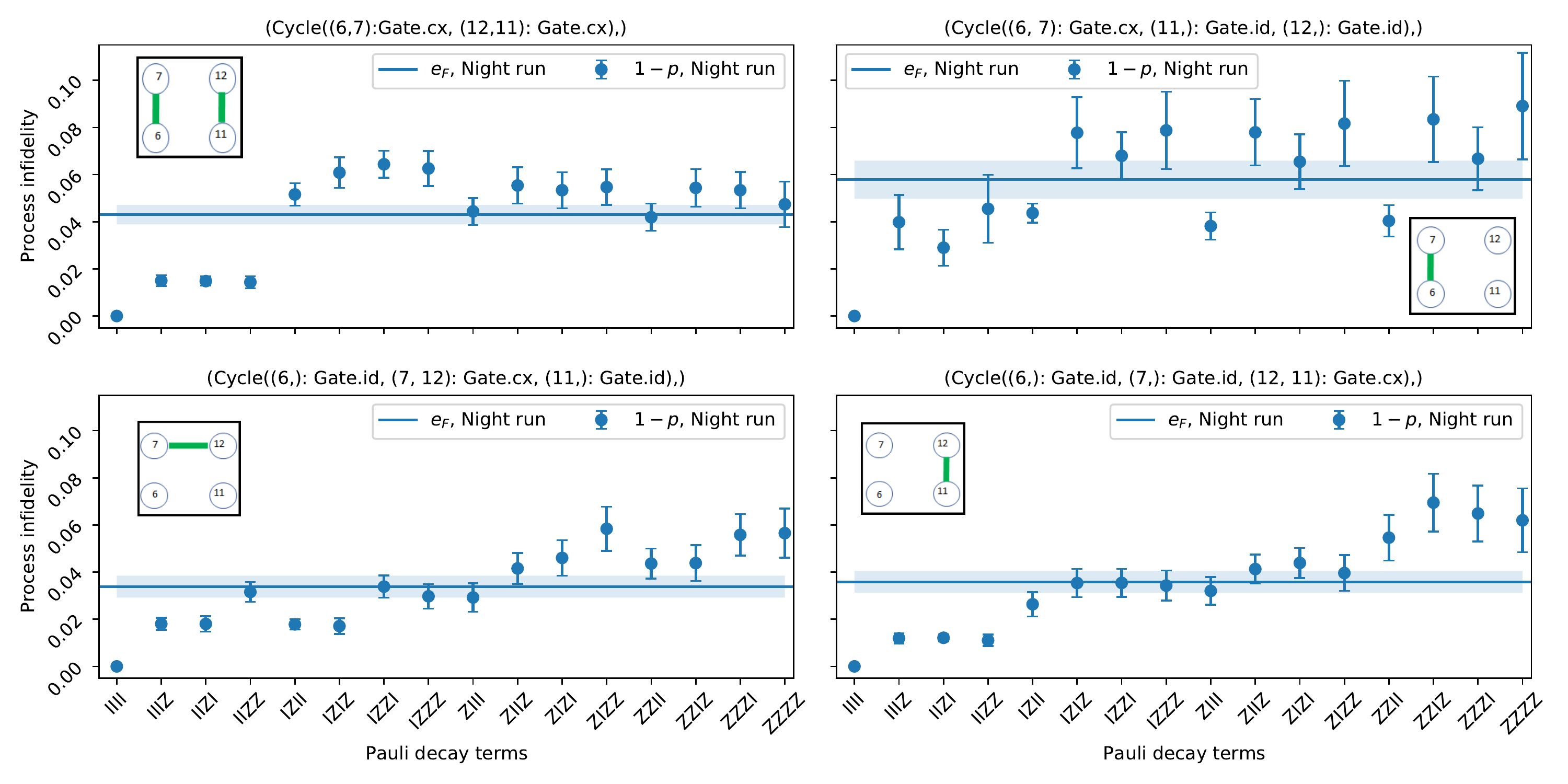}
    \caption{The Pauli infidelities for each hard cycle calculated for each Pauli decay term from night run of layout 2 (qubits [6, 7, 12, 11]) on January 27, 2021. The error bars on each of the markers show the statistical errors on the Pauli decay terms.  The total process infidelity for each of the 4 different cycles are plotted as blue solid lines along with the shaded error bands.  The shaded regions show the error on the average process infidelity}
    \label{fig:Proc_Infid_L2_Night_1_27}
\end{figure*}

\begin{figure*}[!htpb]
    % \centering
    % \includegraphics[width=2.2\columnwidth]{final_plot.pdf}
    \includegraphics[scale=0.51]{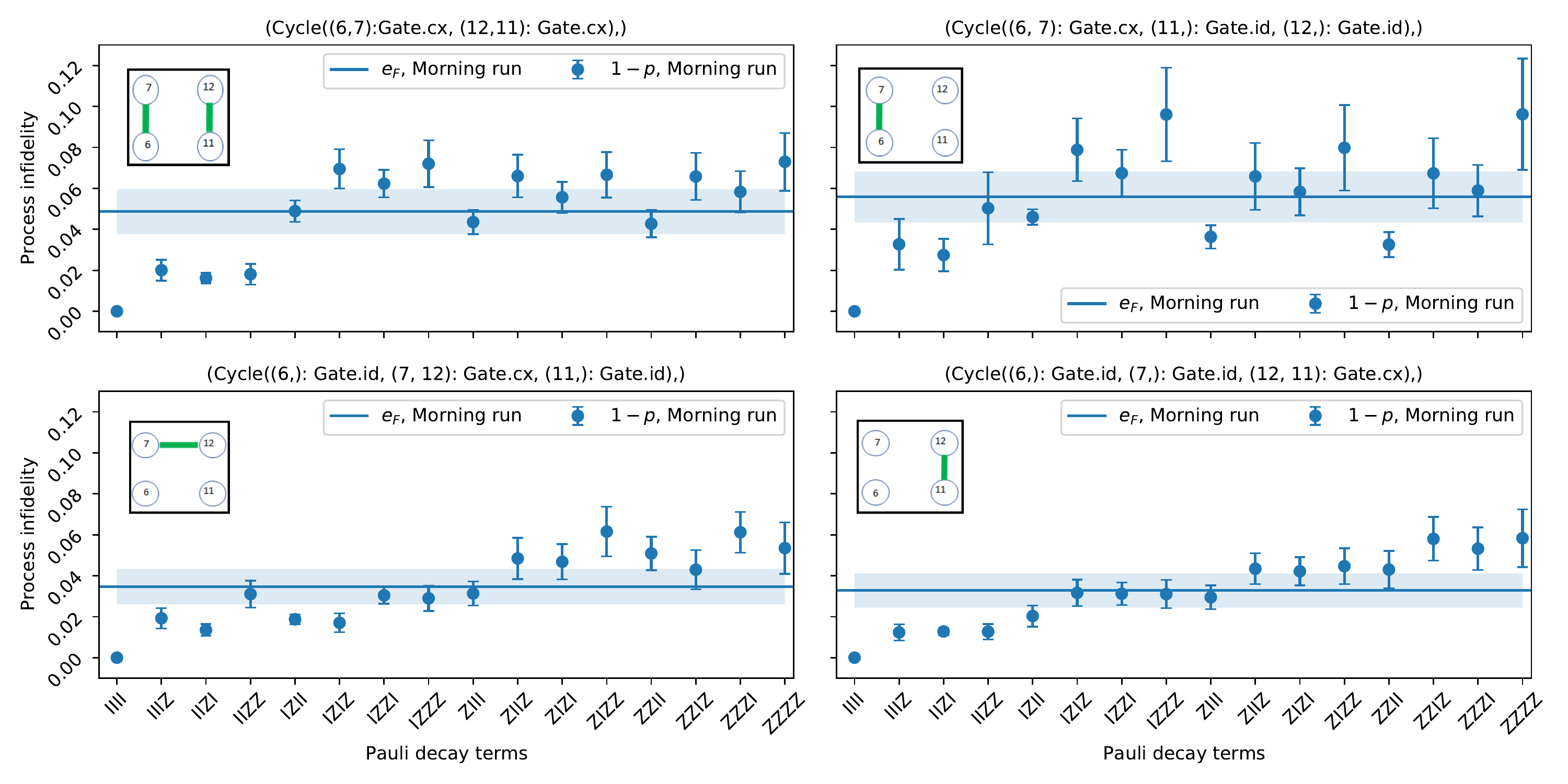}
    \caption{The Pauli infidelities for each hard cycle calculated for each Pauli decay term from morning run of layout 2 (qubits [6, 7, 12, 11]) on 01/30/2021. The error bars on each of the markers show the statistical errors on the Pauli decay terms.  The total process infidelity for each of the 4 different cycles are plotted as blue solid lines along with the shaded error bands.  The shaded regions show the error on the average process infidelity.}
    \label{fig:Proc_Infid_L2_Morning_1_30}
\end{figure*}

%=====================================

\subsection{Analysis of Inter-day and Intra-day Calibration Drift}
\label{sec:inter-day-analysis}

We observed both inter-day and intra-day calibration drifts on the \texttt{ibmq\_boeblingen} processor.  The analysis of the inter-day data shows that both the RB and CB measurements indicate that the processor is substantially drifting from day to day despite both the daily IBM RB re-calibrations and even intra-day (morning and night) IBM re-calibrations of the processor.  This inability of the processor to accurately and faithfully reproduce results within error bars from day-to-day measurements or even measurements recorded at different times within the same day on a totally quiet machine that is free from interference from other users has serious implications as to how users process and interpret their results from quantum computing computations run on these hardware platforms.  Some of these implications include:

\begin{itemize}
    \item Users cannot necessarily assume that the results from applications run on the same qubits on the same processor at the same time each day but on different days can simply be combined together.  A similar concern is raised even when trying to aggregate results run at different times on the same day.  Data recorded when the processor is drifting at these levels have potential statistics implications for users who try to simply aggregate their results from different runs and perform statistical fits on their aggregated data.
    \item The daily IBM one and two qubit re-calibrations of the processor consistently under reported the total magnitude of the qubit process infidelities.  This under-reporting using RB data inflates the magnitude of the signal to noise ratio for data recorded on these processors.  This leads to over-estimates of the total circuit coherence times and has direct implications for the signal to noise ratio justifying where to place the cutoff for the total number of measurements that can be included in a user's data analysis. 
    \item As the QCAP values increase from zero to one, it represents a deterioration in the ability of the circuit to faithfully produce a correct set of measurements at each Trotter step when run on this specific quantum computing hardware platform at two different dates and times. A QCAP value of greater than $0.5$ is usually a qualitative indicator of the deterioration of a processor to faithfully and consistently reproduce results from a specific circuit versus circuit depth (number of Trotter steps). 
    \begin{enumerate}
        \item The two-qubit gates in a circuit are the main source of the errors in the output data.  In terms of circuit depth (number of Trotter steps), the QCAP results using randomized benchmarking for the process infidelities show a far more optimistic scenario as to the number of data points that can actually be justified for inclusion in the user's data analysis as compared to the number of data points that can be justified based on the $QCAP_{CB}$ results.  This has potential implications for analysis of the user's data and implications as to the accuracy of the results.
        \item QCAP graphs are a measurement of the stability of the processor over a time interval. Comparison of QCAP bounds measured at different time intervals from applications run on the same qubits on the same processor at the same time each day but on different days or at different times on the same day should have similar results that fall within each graph's error bars.  Figure~\ref{fig:interdaydrift} and Figure~\ref{fig:intraday_CB_vs_RB} clearly show that this is not the case and consequently results from these measurements cannot be simply combined because that processor is not even stable against drift in the time interval when these measurement were recorded. 
    \end{enumerate}
\end{itemize}

%=====================================

\subsection{Qubit Dependencies for Concurrent Calculations}
\label{sec:qubit choice}

Dependencies on qubit selection and their impact on concurrent calculations on a quantum processor were also investigated.  To generate the data for these measurements identical copies of Circuit 1 were loaded onto qubits located in different physical areas of the hardware platform (Layout 1 and Layout 2) on a specific date and time during the consecutive 8 day running period.  

\begin{figure*}[!htpb]
     \centering
     \includegraphics[scale=0.55,
     clip=true,trim=10 10 10 10]{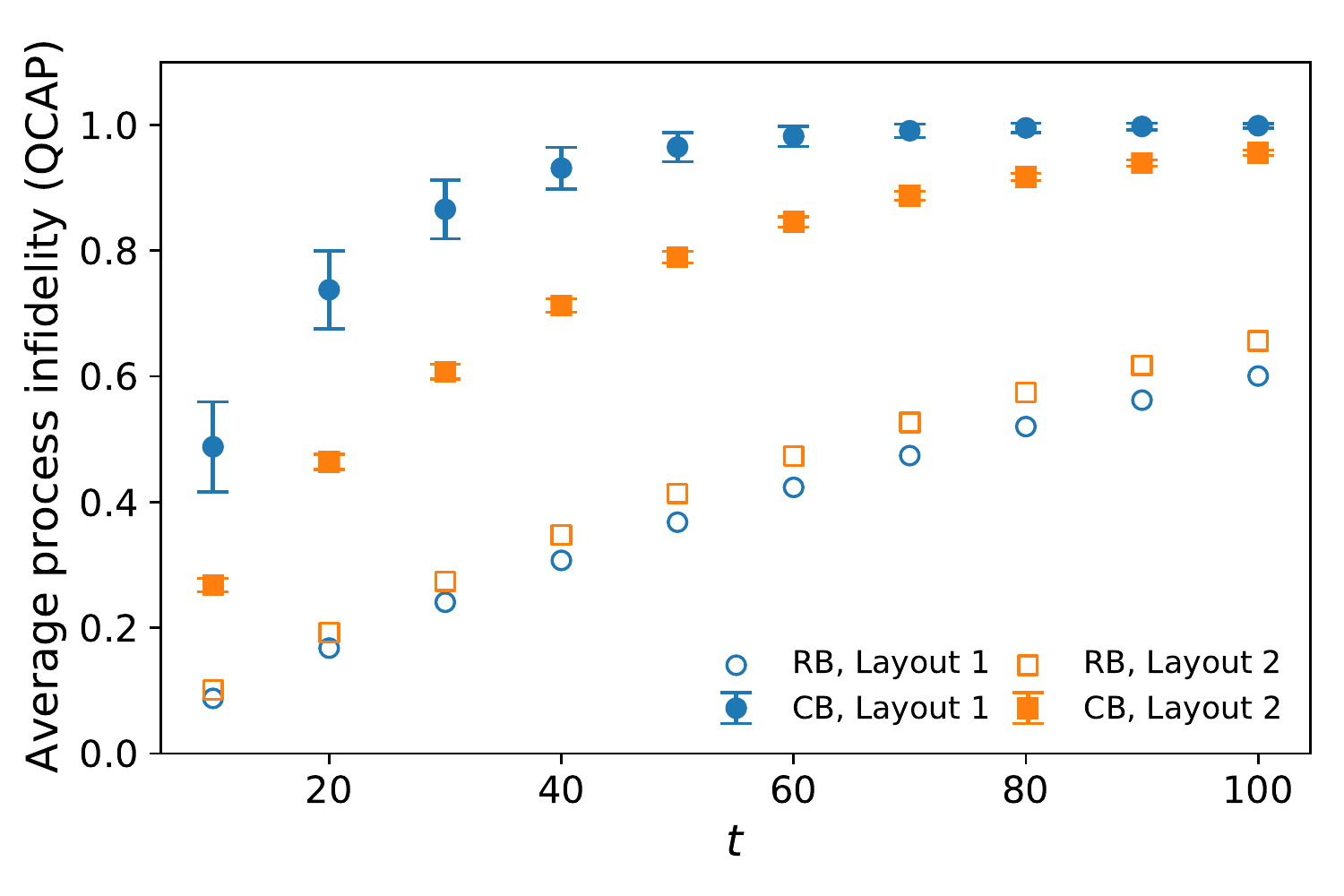} 
     %{QCAP_25Night_27Morning_L1_L2.pdf} 
    \caption{The figure shows the QCAP bound as a function of evolution time calculated using randomized benchmarking (QCAP$_{\textrm{RB}}$) and cycle benchmarking (QCAP$_{\textrm{CB}}$) from the morning run of circuit 1 on both layout 1 and layout 2 on January 27, 2021.}
    \label{fig:QCAP_CB_RB_Data_01_25_night_and_01_27_2021_morning_Layout_1_2C1}
\end{figure*}

An example showing this spatial qubit dependency can be seen from an examination of the January 27$^{th}$ morning run data for Circuit 1 on both Layout 1 (qubits [0, 1, 2, 3]) and Layout 2 (qubits [6, 7, 12, 11]).  Fig.~\ref{fig:QCAP_CB_RB_Data_01_25_night_and_01_27_2021_morning_Layout_1_2C1} shows the QCAP bound as a function of evolution time calculated using randomized benchmarking (QCAP$_{\textrm{RB}}$) and cycle benchmarking (QCAP$_{\textrm{CB}}$) from the morning run of circuit 1 on both layout 1 and layout 2 on January 27, 2021.

Several observations can be deduced from these graphs. 
\begin{enumerate}
     \item  The error bars from the Layout 1 and Layout 2 QCAP$_{\textrm{CB}}$ measurements do not overlap, indicating the reproducibility of Circuit 1 outputs run on different subsets of qubits in non-overlapping areas on the processor are not fully consistent.  The figures also show the length of the time evolution that can indicate valid circuit measurements varies depending on the subset of qubits chosen to run the circuit.
     \item QCAP values greater than $.5$ indicate that the circuit results are impacted by noise and that the results from the computations are less reliable. The QCAP$_{\textrm{RB}}$ shows a substantial difference when compared to QCAP$_{\textrm{CB}}$ measurements. QCAP$_{\textrm{RB}}$ will overestimate the length of the evolution time that the circuit will be able to produce useful data as compared to the length of time indicated from the QCAP$_{\textrm{CB}}$ measurements. 
     \item The QCAP$_{\textrm{RB}}$ measurements are not sensitive to the impacts of coherent errors and crosstalk affecting the results from Circuit 1 running on the different qubit Layouts.
\end{enumerate}

These results indicate that simple parallel processing procedures borrowed from digital computing methods cannot be simply translated and implemented on today NISQ based quantum hardware platforms.  The data clearly show that the individual circuit computations run on Layout 1 and Layout 2 cannot be directly or simply aggregated together because their results cannot be duplicated within the QCAP bound error bars for the computations from each circuit.

\subsection{Circuit Structure Dependencies}
\label{sec:circ1_vs_circ2_QCAP}

A final set of experiments was performed to test the idea that the accumulation of coherent errors strongly depends on the details of the specific circuit layout  being implemented.  To demonstrate this effect we selected two equivalent sets of CNOT gates as shown in Fig.~\ref{fig:circuit_design} (Circuit 1 and Circuit 2).   Although both circuits equivalently represent the TFIM Hamiltonian, the number of steps per cycle for circuit 2 is greater than that for circuit 1 due to the difference in the layout of the CNOT gates. 

\begin{figure}[ht]
    \centering
    \includegraphics[width=0.65\textwidth,
     clip=true,trim=5 15 0 0]{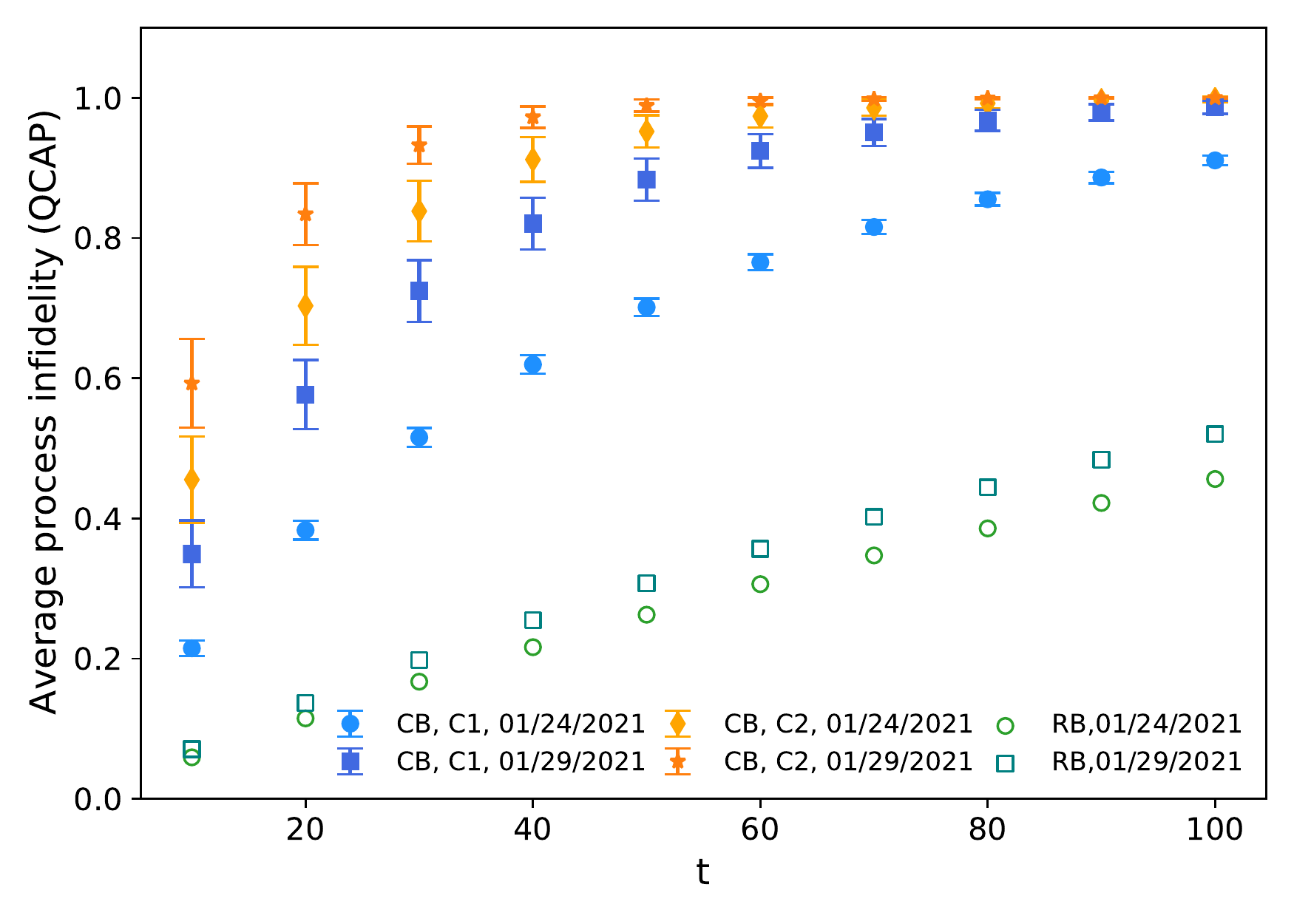}
    \caption{{\bf{Comparing two Trotter decomposition that gives the same physics.}} The QCAP bound as a function of evolution time calculated using RB (QCAP$_{\textrm{RB}}$) and CB (QCAP$_{\textrm{CB}}$) from morning run of layout 2 (qubits [6, 7, 12, 11]) on days 01/24/2021 and 01/29/2021 for Circuit 1 and Circuit 2 with only CNOT gates used as hard cycles. The  plotted  error  bars  only show the statistical error.}
    \label{fig:depth}
\end{figure}

The purpose of using two different layouts that produce the same physics was to investigate the quantum circuit structure dependencies. Data from the morning runs of layout 2 (qubits [6, 7, 12, 11]) on days 01/24/2021 and 01/29/2021 for Circuit 1 and Circuit 2 were selected for analysis.  
The CB QCAP$_{\textrm{CB}}$ bound as a function of evolution time was calculated using only the set of CNOT gates for Circuit 1 and Circuit 2.  The QCAP$_{\textrm{RB}}$ was computed based on the two-qubit backend properties published after the completion of the IBM re-calibrations.  The measured values for both the QCAP$_{\textrm{CB}}$ and QCAP$_{\textrm{RB}}$ bounds for the CNOT hard cycles in circuit 1 and circuit 2 are plotted as a function of evolution time in Fig.~\ref{fig:depth}.  

These results are interesting when comparing these two different types of measurements.  For both the January $24^{th}$ and $29^{th}$ morning run  for Layout 2 data, the QCAP$_{\textrm{CB}}$ bound increases more rapidly than the QCAP$_{\textrm{RB}}$.  There are 3 observations that can be deduced from examining Fig.~\ref{fig:depth}.

\begin{enumerate}
     \item The QCAP$_{\textrm{RB}}$ gives a far too optimistic indication of the time evolution that can be considered for accumulating valid circuit measurements when compared to QCAP$_{\textrm{CB}}$.  This type of result would not be detected if only considering individual CNOT gates and measuring their process infidelity using RB because such a result only depends on the number of CNOT gates and not on the details of the CNOT gate layout itself.  The individual CNOT gates measured through randomized benchmarking QCAP$_{\textrm{RB}}$ calculated from Eq.~\eqref{eq:N_CNOT_error_rates} do not depend on the CNOT layout in the circuit, i.e. they only depend on the number of CNOT gates in the circuit. CB measurements are a better indicator to show how the ordering of the gates in the circuit impact the overall coherence times.
     \item For CB results, the figure shows that the QCAP bound for circuit 2 deteriorates faster than circuit 1 because circuit 2 has a greater number of CNOTs for each cycle.  This is a totally expected result due to the longer depth of Circuit 2 compared to Circuit 1. 
     \item When comparing the QCAP$_{\textrm{CB}}$ measurements between morning runs on Layout 2 for January $24^{th}$ and $29^{th}$ the error bars on these values do not overlap indicating that there is an inter-day CNOT qubit drift occurring and that the measured results from each Circuit are not within statistical consistency from one day to the next.
\end{enumerate}

\section{Observations and Next Steps}
\label{sec:summary_conc}

The results from the computations and measurements reported here illustrate the fragility of today's NISQ platforms.  Several observations from our work here can be readily identified.

The pattern of errors documented in this research are arising from an inability to sufficiently control the coherent and systematic errors from each individual quantum circuit that arises when running on a specific quantum hardware platform.  The two-qubit quantum gates (such as the CNOT) produce error rates that are usually an order of magnitude larger than single qubit gates.  

The coherent errors arising from the under or over rotation of the qubits in response to control pulses significantly contribute to the total qubit error.  This complexity in characterizing the qubit hardware errors indicates that it is not possible to characterize multi-qubit errors by a single parameter.  These hardware error rates on a quantum processor are a complex mix of the choice of qubits and choice of gates, the order or sequence that the gates are being applied in the circuit and the local direct gate combinations and surrounding spectator qubits.

Consequences of these coherent errors appear when running benchmark type computations within a controlled hardware environment.  These instabilities appear as both inter-day and intra-day qubit drift compromising the consistency of quantum computation results as well as physical qubit idiosyncrasies dependent on which physical susbset set of qubits is used for the computations.  

At the present time the number of qubits used for these computations is quite modest and so the accuracy of the quantum computing results can always be checked with digital computers.  However, as the number of qubits used in quantum computations over the next several years exceeds a few hundred, the hardware will still not have a sufficient number of qubits to support logical error correction for these computations.  In addition, a direct verification of the quantum computer's output will be far beyond the reach of any digital computer.

If the user community is to have any level of confidence in the published results from running quantum circuits using several hundred qubits, there must be a demonstrated ability to control the noise and errors on these systems.  These mitigation strategies must demonstrate that the measurements from applications run on these quantum hardware platforms are sufficiently stable so that the results can be trusted with a high level of confidence.  Without such supporting evidence, these computations will be overshadowed by the possibility that the data is tainted with some of the instabilities as documented in this report and will be considered unreliable.

We recognize that \texttt{ibmq\_boeblingen} is an older generation quantum processor.  IBM is well aware of these noise and coherent error issues and they have a large complex hardware and software effort underway to address these challenges~\cite{PhysRevLett.119.180509},\cite{ mckay2020correlated}, \cite{ 2021_Bravyi}, \cite{ ware2019crossresonance}, \cite{ 2021_Nation}, \cite{ kim2021scalable}.  Although there has been progress with both the hardware and software since the release of the \texttt{ibmq\_boeblingen} processor, the error mitigation protocols and methodologies summarized here are still applicable. 

IBM has also recently introduced the IBM Qiskit Runtime Program to help streamline computations and minimize intra-day drift.  In addition, other groups are investigating additional methods for improving the reliability of quantum processing computations~\cite{PhysRevResearch.3.033285, 2021,zen,zen2,mdn} and comparisons among methods would be desirable. 

This group has also begun several other in-depth follow-on projects to address other aspects concerning the integrity and validity of NISQ computations through robust characterization and stability analysis protocols. These include a detailed analysis of the impacts of coherent errors on magnon spectra measurements and real-time phase shifts. The results from these ongoing projects will be reported in future publications.

\section{Acknowledgments}

PD was supported in part by the U.S. Department of Energy (DoE) under award DE-AC05-00OR22725.  KYA and RCP were supported by the Quantum Information Science Enabled Discovery (QuantISED) for High Energy Physics program at ORNL under FWP number ERKAP61 and used resources of Oak Ridge Leadership Computing Facility located at ORNL, which is supported by the Office of Science of the Department of Energy under contract No. DE-AC05-00OR22725.  YM and EG are supported by a Department of Energy QuantiSED grant DE-SC0019139.  ZP gratefully acknowledges funding support from the NSF with an NSF pre-doctoral fellowship.  We acknowledge the use of IBM Quantum services for this work, especially discussions with Nathan Earnest-Noble, Matthew Stypulkoski, Azia Ngoueya and Patrick Mensac from IBM.  We thank North Carolina State University (NCSU) for access to the IBM Quantum Network quantum computing hardware platforms through the NCSU IBM Quantum Hub and thank IBM Research for the extended dedicated mode reservations and availability of the \texttt{ibmq\_boeblingen} processor on which the computations were performed.  The views expressed are those of the authors and do not reflect the official policy or position of IBM or the IBM Quantum team. The project team acknowledges the use of TrueQ software from Keysight Technologies and useful discussions with Ian Hincks, Dar Dahlen, and Arnaud Carignan-Dugnas from Keysight. 
 
 \section{Author Contributions}
 P.D. led the team, designed the operational implementation, coordinated with IBM Research for dedicated access to the IBM quantum hardware platforms and was a major contributing author for this paper.  KYA ran the cycle benchmarking simulations and constructed the CB and QCAP graphs and made substantive contributions to the text. ZP captured the daily IBM backend property data.  ZP also ran the TFIM simulations and wrote the analysis software for post-processing and plotting the TFIM data. EG provided the orignal TFIM code and the physics model circuits. AN built the tables in the paper. AFK, RP and YM edited and reviewed the document.
 
 \section{Competing interests}
 The authors declare no competing interests
 
 \clearpage

%\bibliographystyle{apsrev4-1}

%==========================================================%
%\onecolumn
\bibliographystyle{iopart-num}
\bibliography{bibliography}
%\bibliography{main.bbl}
%\twocolumn
%==========================================================%

\section{Appendix}
\label{sec:appendix}
\begin{appendix}

\section{Daily \texttt{ibmq\_boeblingen} Qubit Re-Calibration Schedule}
\label{IBM_re_calibrations}

For this specific project IBM agreed to supply our team with approximately 140 hours of dedicated reservation time and to follow an agreed upon customized calibration schedule.

The customized schedule for dedicated time included a period in the morning from 4 am until 10 am and again in the afternoon from 3 pm until 11 pm.  The complete \texttt{ibmq\_boeblingen} re-calibration for both single and two-qubit gate gates was scheduled at 4:00 am ET, the beginning of the morning dedicated reservation time.  A second re-calibration for only two-qubit gates ran at 6:00 pm ET, approximately 3 hours into the afternoon dedicated reservation time. The calibration jobs took approximately an hour and a half to complete. Our team executed no external jobs on the device during the calibration process, allowing the calibration jobs to run without interference.

The single-qubit calibration process consisted of Ramsey and Rabi experiments to measure the frequency and amplitude of each qubit along with calibration of the optimal scaling factor of the Derivative Removal by Adiabatic Gate (DRAG) pulse used in single-qubit gates on superconducting hardware. The T1/T2 coherence times and measurement error rates of each qubit were also measured and recorded. Randomized benchmarking of the single-qubit gates was then performed in batches of non-adjacent qubits. The two-qubit calibration process was done in a similar manner.  Calibration of the amplitude and phase of each pulse was completed before performing randomized benchmarking in batches of well-separated gates of similar length in order to measure the average gate fidelities. Each time the \texttt{ibmq\_boeblingen} quantum computing hardware platform was re-calibrated and benchmarked, IBM published and made these backend properties available through Qiskit, the open-source quantum software development kit.

\section{Cycle Benchmarking and Quantum Capacity (QCAP) Protocols}
\label{Appendix_CBandQCAP}

This appendix summarizes both cycle benchmarking and quantum capacity protocols and their True-Q software implementation and parameter settings used for the computations reported in this paper.

\subsection{Cycle Benchmarking}
\label{Appendix:cyclebenchmarking}

Cycle benchmarking (CB) is a scalable noise characterization protocol that was selected to identify local and global errors across multi-qubit quantum processors.  The CB protocol can measure errors such as process infidelity containing any combination of single gates, two-qubit gates, and idle qubits, across an entire quantum device.  CB helps keep track of each twirling gate and makes the process scalable with the number of qubits~\cite{ville2021leveraging}.

This protocol has the feature that the number of measurements required to estimate the process fidelity to a fixed precision is approximately independent of the number of qubits and is also insensitive to State Preparation and Measurement (SPAM) errors.  Robustness to SPAM is very important characteristic because these type of errors can dominate the gate error measurement.

\begin{figure*}[!htpb]
%\begin{figure}
    \centering
    \includegraphics[width=1.05\textwidth]{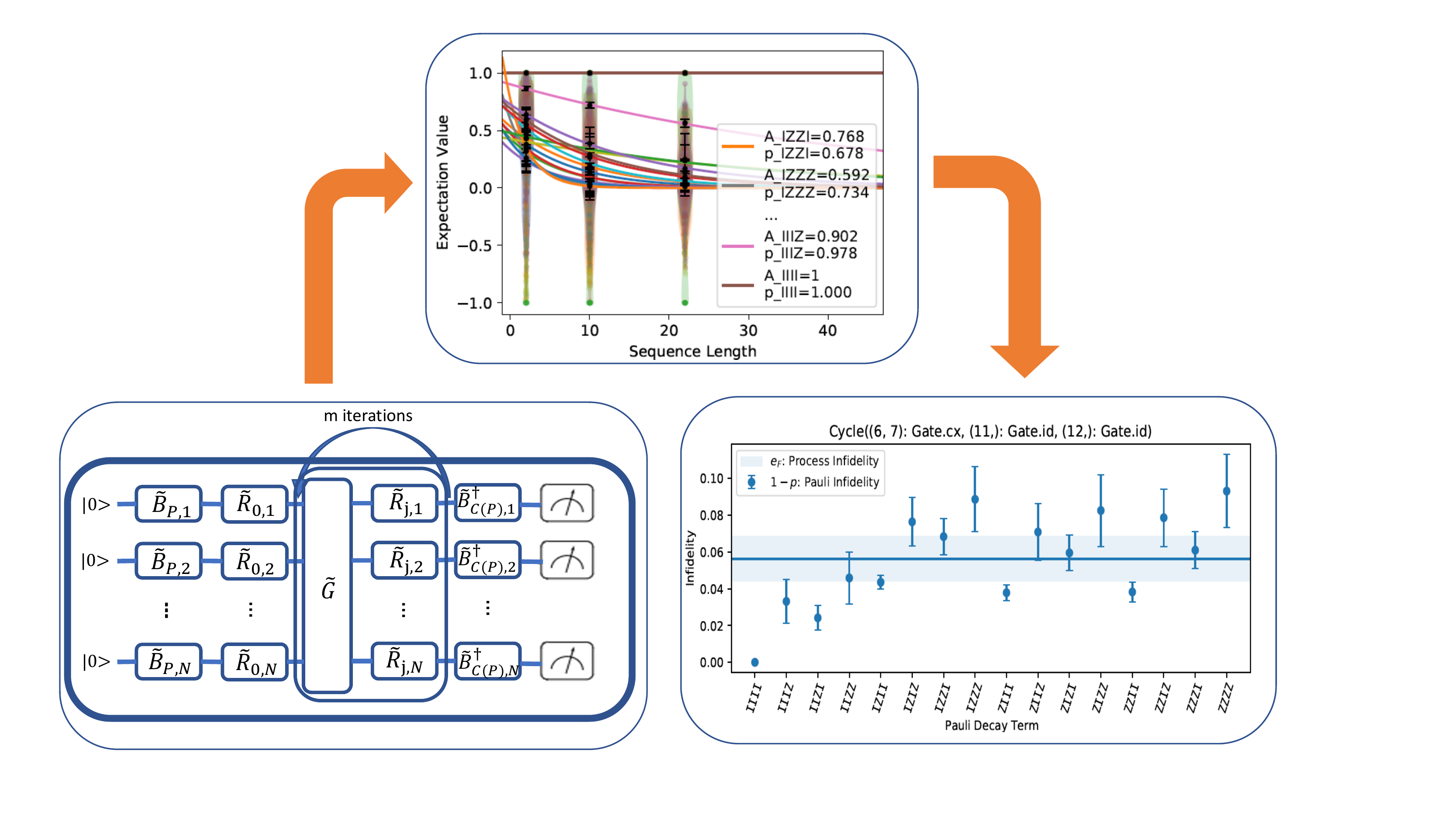}
\caption{Block diagram of the CB protocol implementation.}
\label{fig:cb_diagram_1_2_3}
\end{figure*}

The CB protocol is presented in detail in reference~~\cite{Erhard2019} and is schematically represented in Figure~\ref{fig:cb_diagram_1_2_3}.   In CB, a gate cycle is an arbitrary set of native operations that act on a quantum register within a single clock cycle of time. Furthermore, within the CB protocol, there is a distinction between operations that can be physically implemented with relatively small and large amounts of noise, respectively called `easy' and `hard' gate cycles.

The box on the left hand side of the figure shows the CB protocol ``dressing" a primitive gate cycle of interest ( represented by $\mathcal{\tilde{G}}$ ) by composing the cycle with independent, random n-qubit Pauli operators in such a way that the effective logical circuit remains unchanged.  In the figure the block $\mathcal{\tilde{G}}$ represents the noisy implementation of the gate(s) being measured in the circuit.  The blocks $\mathcal{\tilde{R}}_{i,j}$ are random Paulis represented by the $j^{th}$ tensor factor of the $i^{th}$ gate inserted into the cycle to create an effective Pauli channel for the gate $\mathcal{\tilde{G}}$ being measured.  The blocks $\mathcal{\tilde{B}}$ and $\mathcal{\tilde{B}}^{\dagger}$ represent basis changing operations connected with controlling SPAM errors.

CB decouples state preparation and measurement errors from the process fidelity estimation of a particular gate cycle by applying the noisy, dressed cycle to the system m number of times, (called the sequence length), and extracting the process fidelity from the average decay rate as a function of this sequence length. This Pauli twirling of gate cycles map coherent errors into stochastic Pauli errors, which are then measured in the prepared eigenstates of the Pauli basis set.

This is represented by the top box in the center of the figure showing all measured Pauli decay expectation values plotted as a function of the gate sequence length.  In practice, this process is computed using at least three distinct gate sequence lengths.  Each measurement sequence produces an exponential decay of the expectation value versus the sequence length.  Taken together, the set of exponential decays of the form $Ap^{m}$ can be fit to the cycle of interest as a function of the circuit depth for each basis preparation state.

Using the fitted exponential decay, the individual process infidelity for each Pauli Decay term $e_{F}$ can be measured as shown in the box on the right hand side of the figure.  An average process infidelity and error for the particular cycle $\mathcal{\tilde{G}}$ is calculated and is represented by the solid line and shaded band on the graph.

For our project in order to measure the error characterization associated with the two qubit gates in the TFIM circuits, the cycle benchmarking (CB) protocol was implemented using the True-Q software package.  This package included a function \emph{make\_cb} that can produce quantitative measurements showing the effect of global and local error mechanisms affecting different primitive cycle operations of interest using CB.

The \emph{make\_cb} in True-Q uses a set of input parameters for the calculation.  The first parameter is the cycle of interest. The second parameter sets how many times to apply the dressed cycle to observe the decay of the expectation values. Here, dressed cycle is the term that is used for denoting the target cycle preceded by a cycle of random elements of the twirling group. The number of random cycles needs to be chosen carefully such that it leads to exponential decay is evident and the fidelity can be accurately estimated.

The third parameter in the function is the number of circuits for each circuit length determined in the second parameter, i.e. number of random cycles. The last parameter in the function is the number randomly chosen Pauli decay strings. One can also specify the twirling group to be used that will be used in the process to automatically instantiate a twirl based on the labels in the given cycles. The supported twirling groups in True-Q software are Pauli, Clifford, unitary and identity. The software also offers initializing a twirl with single-qubit Cliffords. After the circuits are generated using this function the expectation values of the Pauli operators are calculated which then gives the process infidelity for the cycle of interest by using an exponential fit to the decay of the expectation values.

The Clifford (C1) gates for the hard gate twirling were selected to minimize the computation time so that they would complete within the morning and night dedicated time windows available on \texttt{ibmq\_boeblingen}. The C1 twirling used random single qubit Cliffords which had the effect of symmeterizing the $X$, $Y$ and $Z$ noise.  This ultimately allowed for an analysis of the depolarization error, which is one of the simplest of the systematic errors to measure and study.

To calculate the contribution of each of the Pauli decay terms to the average process infidelity $e_F$, C1 twirling was done using gate sequence circuit lengths of 2, 10, and 22 random Clifford gates were applied to each of the different pair combinations of CNOTs.  Here, the sequence length refers to the number of times the cycle of interest appears apart from state inversion. We used 48 random circuits in each sequence length, and  128 shots. The combination of the CNOT gate being measured and the sequence of random Cliffords defines a dressed cycle of the CNOT gates being measured.  For each three different circuit lengths the expectation values were calculated for all 16 of the Pauli decay terms.  From these expectation values, fits to the exponential decay $A p^m$ (SPAM parameter $A$ and the decay parameter $p$) are calculated for each Pauli decay term.

Individual process infidelity measurement were recorded for every CNOT pair for each of the three different qubit layouts on \texttt{ibmq\_boeblingen} device as shown in Fig.~\ref{fig:hardware_layout}. For example, on Layout 1 measurements included all of the combination of two-qubit cycles ([0, 1 and 2, 3], [0, 1], [1, 2] and [2, 3]).  Similar measurements were taken on the CNOT cycles for Layouts 2 and 3.  Hence, there are four cycles studied for each qubit layout

The average process infidelity of the dressed cycle for that CNOT pair was computed based on the calculated values of each of the Pauli decay terms.  Both the individual process infidelity and average process infidelity measurements were computed and used in the stability analysis of the qubits on the \texttt{ibmq\_boeblingen} processor.  The individual process infidelities for each CNOT pair and the overal process infidelity are shown in Fig~\ref{fig:PauliInfidelities24_29_Story4}, Fig~\ref{fig:Proc_Infid_L2_Morning_1_27}, Fig~\ref{fig:Proc_Infid_L2_Night_1_27}, Fig~\ref{fig:Proc_Infid_L2_Morning_1_30} and Fig~\ref{fig:PauliInfidelities30Night}.

%===========================================================

%\end{figure}

\subsection{Quantum Capacity Bound}
\label{sec:qcap}

The QCAP protocol was used for comparing the measured performance of a circuit that is loaded onto a quantum computing hardware processor to the measurement of an equivalent idealized version of that same circuit.  A value of ``0" for a QCAP result means that the circuit being tested is identical to its idealized equivalent whereas a QCAP value of``1" implies that the circuit being measured has no equivalent performance characteristics to its idealized equivalent.  An increase in the QCAP bound as a function of evolution time is a measure as to how many time evolution steps can be included in a result before the signal being measured is overcome by noise in the circuit.

For the actual QCAP measurements the \emph{make\_qcap} and \emph{qcap\_bound} functions in the True-Q software were used to obtain an equivalent bound on the performance of a circuit as if it were computed using randomized compiling for calculating the process infidelity of the entire circuit of interest.  The parameters to generate the collection of circuits for the quantum capacity bound \emph{make\_qcap} function use similar parameters as \emph{make\_cb} function, i.e. the circuit of interest, a list for the number of random cycles, number of circuits for each random cycle, and total number of randomly chosen Pauli decay strings. After generation of quantum circuits, these circuits are embedded into \emph{qcap\_bound} as well as the circuit of interest to return a bound on the circuit performance.  In this particular project, due to limited access to the dedicated mode on \texttt{ibmq\_boeblingen} device we used sequence lengths of 4 and 16. The number of random circuits in this case is 30 and each of these circuits were run $N_{\text{shots}}=128$.

We selected three separate groups of qubits on the \texttt{ibmq\_boeblingen} hardware platform as shown in Fig.~\ref{fig:hardware_layout} to study the error characterization of TFIM Trotterization circuits using CB.  The quantum circuit for evolution under the TFIM Hamiltonian has three pairs of two-qubit CNOT gates (c.f. Fig.~\ref{fig:circuit_design}).  For the QCAP computation we selected circuit 1 shown in Fig~\ref{fig:circuit_design} in order to compare to previous TFIM measurements \cite{Gustafson2021}.  We computed an estimate to the QCAP bound of the circuit 1 CNOT cycles in the TFIM Trotterization quantum circuits as a function of the number of Trotter steps.  We also calculated the QCAP bound from the CNOT error rates reported by IBM using RB.  To this end, we used the expression for the relationship between the average process fidelity and the average gate fidelity as seen in Eq.~\eqref{eq:err_rate}. For a quantum circuit with $N$ CNOT gates (Eq.~\ref{eq:N_CNOT_error_rates}) the QCAP bound is calculated using CNOT error rates provided by IBM.
\begin{equation}
    \text{QCAP}_{\text{RB}}=1-\prod_{i=1}^N\left(1-\frac{d+1}{d} r_i\right)~.
    \label{eq:N_CNOT_error_rates}
\end{equation}
The QCAP bound versus step size was then plotted as the average process infidelity (QCAP bound) as a function of number of Trotter steps (as a function of time).  From this graph the performance of the circuit implemented on the set of specific qubits on that specific hardware platform can be measured over time.

\clearpage

\section{Tables}
\label{Appendix:Tables}

\begin{table*}[h]
\begin{tabular}{|c|c|c|c|c|c|c|}
\hline \hline
\multicolumn{7}{|c|}{\textbf{Inter-Day, Layout 2, Single-Qubit Errors}}   \\ \hline \hline
  & \textbf{Qubits} & $\mathbf{T_1 (\mu s )}$  & $\mathbf{T_2 (\mu s )}$ & \textbf{\begin{tabular}[c]{@{}c@{}}IBM Backend\\ Readout Error\\
  ($\times10^{-2}$) \end{tabular}} & \textbf{\begin{tabular}[c]{@{}c@{}}$\mathbf{U_2}$\\ ($\times10^{-4}$)\end{tabular}} & \textbf{\begin{tabular}[c]{@{}c@{}}$\mathbf{U_3}$\\ ($\times10^{-4}$)\end{tabular}} \\ \hline
\multirow{4}{*}{\begin{tabular}[c]{@{}c@{}}01/24/2021\\ Morning Run\end{tabular}} & 6   & 67.1  & 99.9  & 2.54  & 2.87  & 5.74  \\ \cline{2-7}
                                                                                  & 7   & 94.8  & 86.8  & 2.30  & 3.05  & 6.71  \\ \cline{2-7}
                                                                                  & 12  & 97.5  & 88.5  & 3.13  & 2.91  & 5.82  \\ \cline{2-7}
                                                                                  & 11  & 95.1  & 71.6  & 3.55  & 4.44  & 8.88   \\ \hline
\multirow{4}{*}{\begin{tabular}[c]{@{}c@{}}01/29/2021\\ Morning Run\end{tabular}} & 6   &24.1  & 4.97   & 9.19  & 25.0  & 50.0   \\ \cline{2-7}
                                                                                  & 7   & 78.8  & 103.4  & 2.84  & 3.97 & 7.94   \\ \cline{2-7}
                                                                                  & 12  & 80.8 & 114.6  & 3.47  & 3.49  & 6.98   \\ \cline{2-7}
                                                                                  & 11  & 48.4 & 87.8  & 3.22  & 4.64   & 9.29  \\ \hline
\end{tabular}

\caption{$T_1$, $T_2$ values, readout errors and single-qubit errors for basis gates $U_2$ and $U_3$ for Layout 2 (qubits [6,7,12,11]) extracted from the recorded IBM back-end properties immediately after IBM completed a full re-calibration of the \texttt{ibmq\_boeblingen} quantum chip on the morning of January 24, 2021 and January 29, 2021.}
\label{table:Single-qubit-gate-error-layout2-on-1-24-and-1-29}
\end{table*}

\begin{table*}[h]
\vspace{0.25in}
\begin{center}
\begin{tabular}{|c|c|c|c|c|}
\hline \hline

\multicolumn{5}{|c|}{{\textbf{Inter-Day, Layout 2, Two-Qubit Process Infidelities}}}

\\ \hline \hline
  & \textbf{Cycle}    & \textbf{Qubits} & \textbf{\begin{tabular}[c]{@{}c@{}}Cycle Benchmark\\ $(\times 10^{-2})$\end{tabular}} & \textbf{\begin{tabular}[c]{@{}c@{}}IBM Backend\\ $(\times 10^{-2})$\end{tabular}} \\ \hline
  & 2  & [6,7]  & 3.67  & 0.908   \\ \cline{2-5}
  & 3   & [7,12]   & 2.08  & 1.128  \\ \cline{2-5}
\multirow{-3}{*}{\begin{tabular}[c]{@{}c@{}}01/24/2021\\ Morning Run\end{tabular}} & 4  & [12,11]  & 2.07  & 1.010   \\ \hline
  &  2 & [6,7]  & 4.79 & 3.30  \\ \cline{2-5}
  &  3 & [7,12]  & 3.34  & 1.25  \\ \cline{2-5}
\multirow{-3}{*}{\begin{tabular}[c]{@{}c@{}}01/29/2021\\ Morning Run\end{tabular}} & 4 & [12,11] & 3.52  & 1.124  \\ \hline
\end{tabular}
\end{center}
\caption{Values for two-qubit process infidelities for qubit pairs [6, 7], [7,12], [12, 11] extracted from the recorded IBM back-end properties immediately after IBM completed a full re-calibration of the \texttt{ibmq\_boeblingen} quantum chip on the morning of January 24, 2021 and January 29, 2021 and the process infidelity values for the two-qubit pairs from cycle 2, 3, and 4 from the cycle benchmarking computations.}
\label{table:Two-qubit-gate-error-layout2-on-1-24-and-1-29}
\end{table*}

\begin{table*}[h]
\vspace{0.25in}
\begin{center}
\label{table:Intraday_2_Qubit_Proc_Infidelities_27_30}
\begin{tabular}{| c | c | c | c | c |}
\hline \hline
\multicolumn{5}{|c|}{{\textbf{Layout 2 Intra-Day Two-Qubit Process Infidelities}}}   \\ \hline \hline
   & \textbf{Cycle} & \textbf{Qubits} & \textbf{\begin{tabular}[c]{@{}c@{}}Cycle Benchmark\\  $(\times 10^{-2})$\end{tabular}} & \textbf{\begin{tabular}[c]{@{}c@{}}IBM Backend\\ $(\times 10^{-2})$\end{tabular}} \\ \hline
   & 2     & [6,7]   & 5.45   & 3.23    \\ \cline{2-5}
  & 3     & [7,12]  & 3.64    & 1.08    \\ \cline{2-5}
\multirow{-3}{*}{\begin{tabular}[c]{@{}c@{}}01/27/2021\\ Morning Run\end{tabular}} & 4    & [12,11]   & 2.77  & 0.98   \\ \hline
 & 2   & [6,7]  & 5.78  & 3.38   \\ \cline{2-5}
 & 3   & [7,12]   & 3.38  & 1.11  \\ \cline{2-5}
\multirow{-3}{*}{\begin{tabular}[c]{@{}c@{}}01/27/2021\\ Night Run\end{tabular}}  & 4  & [12,11]  & 3.59  &  1.12   \\ \hline
  & 2   & [6,7]  & 5.59   & 2.85   \\ \cline{2-5}
  & 3   & [7,12]  & 3.47  & 1.08   \\ \cline{2-5}
\multirow{-3}{*}{\begin{tabular}[c]{@{}c@{}}01/30/2021\\ Morning Run\end{tabular}} & 4 & [12,11]  & 3.52  & 1.06  \\ \hline
  & 2   & [6,7]   & 4.86  & 3.64 \\ \cline{2-5}
  & 3   & [7,12]  & 3.63  & 1.40  \\ \cline{2-5}
\multirow{-3}{*}{\begin{tabular}[c]{@{}c@{}}01/30/2021\\ Night Run\end{tabular}}   & 4  & [12,11]  & 3.28  & 1.09   \\ \hline
\end{tabular}
\end{center}
\caption{Values for two-qubit process infidelities for qubit pairs [6, 7], [7,12], [12, 11] extracted from the recorded IBM back-end properties immediately after IBM completed full re-calibrations of the \texttt{ibmq\_boeblingen} quantum chip on the morning of January 27th and 30th 2021 and after the two-qubit re-calibrations at night on January 27th and 30th 2021. The table also lists the process infidelities for the two-qubit pairs from cycle 2, 3, and 4 obtained from the cycle benchmarking computations for those same time periods.}
\label{table:Two-qubit-gate-error-layout2-on-1-27-and-1-30}
\end{table*}

\end{appendix} 

\end{document}